\documentclass[aps,prb,twocolumn,showpacs,superscriptaddress,groupedaddress]{revtex4-1}
\usepackage{graphicx}
\usepackage{amsmath}
\usepackage{amssymb}
\usepackage{bm}
\usepackage{color}
\usepackage{orcidlink}
\usepackage[ruled]{algorithm2e}
\usepackage{xcolor} 
\colorlet{RED}{red} 
\newcommand{\w}[1]{\textcolor{black}{#1}}
\newcommand{\rev}[1]{{\color{black}#1}}

\SetKwComment{Comment}{// }{}

\SetCommentSty{mycommfont}

\begin{document}

\title{Emergent Wigner-Dyson Statistics and \rev{Self-Attention-Inspired} Prediction in Driven Bose-Hubbard Chains}

		\author{Chen-Huan Wu 
			\orcidlink{0000-0003-1020-5977} }
		\thanks{chenhuanwu1@gmail.com}
                \affiliation{Department of Physics, Faculty of Science, Universiti Malaya, Kuala Lumpur 50603, Malaysia}

\begin{abstract}
	We propose an algorithm based on modulable hidden variables and adaptive step lengths, inspired by  statistical physics and renormalization group principles, to study the emergent Wigner-Dyson level statistics in a driven many-body system.
	We apply this method to the boundary-driven Bose-Hubbard chain to illustrate the competition between coherent driving, hopping, and on-site interactions.
Unlike random matrix models where chaos is induced through quenched disorder, here ergodicity emerges intrinsically from many-body configuration hybridization induced by the interplay between the driving field $F$ and the nonlinearity $U$.
	The resulting system follows 
    statistics approaching the Gaussian orthogonal ensemble (GOE) limit.
\rev{Using a self-attention-inspired Gaussian relevance map, the algorithm organizes the discrete site-occupation features and iteratively redistributes a coarse-grained spectral measure. This reduces the computational overhead relative to calculating the full eigenspectrum, although the truncated Hilbert-space dimension $D$ itself still grows exponentially with system size. The deterministic feedback stabilizes the macroscopic spectral variance, while a separate stochastic step restores the local fluctuations required for a GOE-like surrogate spacing distribution. The reconstruction loop therefore does not require a full eigendecomposition, while exact diagonalization is retained here only for finite-size validation.}
\end{abstract}

\maketitle

\section{Introduction}

\rev{The study of thermalization and chaos in isolated quantum many-body systems has drawn significant attention\cite{Deutsch,Srednicki,Rigol,Mehta,Ferrari,Solanki,Neuenhahn}.}
\rev{We consider a one-dimensional (1D) driven Bose-Hubbard lattice of length $L$, indexed by $\ell$ ($\ell=1,\cdots,L$).}
While standard integrability breaking is often associated with random disorder, here we investigate the intrinsic chaos generated by the interplay of strong on-site interaction and a coherent driving field at the boundary.


The emergence of many-body chaos requires sufficiently strong hybridization in Hilbert space together with interactions that generate level repulsion. To characterize this competition, we introduce a reconstruction framework based on modulable hidden variables, where the diagonal interaction energies serve as the underlying hidden manifold and the spectral weights are determined self-consistently through thermodynamic feedback.
\rev{The practical advantage is that the reconstruction updates a coarse-grained probability measure rather than computing the full many-body eigenspectrum. In particular, the target second moment can be evaluated from the trace invariants $\mathrm{Tr}(H)$ and $\mathrm{Tr}(H^2)$ without diagonalizing $H$; exact diagonalization is used below only as an independent finite-size benchmark for the adjacent-gap ratio and eigenstate IPR.}
\rev{This reduction should not be interpreted as eliminating the exponential growth of the truncated Hilbert space: $D=(N_{\max}+1)^L$ still increases exponentially with $L$. Rather, the present construction avoids a full eigendecomposition inside the reconstruction loop and replaces it by trace/moment evaluation and iterative updates of $D$ spectral weights. The numerical validation in this work is deliberately restricted to $L=4$--$8$, where ED remains feasible and can serve as an independent benchmark.}

\rev{Throughout the manuscript, ``self-attention-inspired'' refers to a physically constrained, distance-based relevance mapping acting on Fock-configuration features. In the implemented reduced form, the many-body configurations are mapped to the scalar interaction-energy features $M_i$, and their relevance is evaluated relative to the dynamically aggregated weighted mean $\mu_t=\sum_j w_jM_j$. The projection matrices of the generic Query--Key--Value notation are therefore not independently trained by a conventional Transformer loss; the optimized dynamical variables in the present implementation are the spectral weights $w_i$.}
\rev{A complementary recent analysis of a bosonic many-body model emphasized the usefulness of combining unfolded level statistics with consecutive gap ratios and eigenstate diagnostics when distinguishing integrable and chaotic regimes, particularly in the presence of finite-size effects and unfolding ambiguities\cite{Prashant2026}. This motivates our use of the exact adjacent-gap ratio as an independent benchmark for the reconstructed surrogate statistics.}

The Hamiltonian describing this driven Bose-Hubbard system reads\cite{Ferrari,Solanki}
\begin{equation}
	\begin{aligned}
		\label{Hb}
		H_{b} = & \sum_{\rev{\ell}=1}^{L} \left( \Delta \hat{a}^\dagger_{\rev{\ell}} \hat{a}_{\rev{\ell}} + \frac{1}{2} U \hat{a}^\dagger_{\rev{\ell}} \hat{a}^\dagger_{\rev{\ell}} \hat{a}_{\rev{\ell}} \hat{a}_{\rev{\ell}} \right) \\
		& - \sum_{\rev{\ell}=1}^{L-1} J \left(\hat{a}^\dagger_{\rev{\ell}+1} \hat{a}_{\rev{\ell}} + \hat{a}^\dagger_{\rev{\ell}} \hat{a}_{\rev{\ell}+1}\right) + F(\hat{a}^\dagger_1 + \hat{a}_1),
	\end{aligned}
\end{equation}
where $\Delta$ is the detuning, $U$ is the on-site interaction strength, $J$ is the hopping amplitude, and $F$ represents the coherent driving field applied to the first site. \rev{Throughout the numerical calculations presented in this work, we set the detuning to $\Delta=0$.}
\rev{Here, the diagonal interaction term $\frac{U}{2}\hat n_{\ell}(\hat n_{\ell}-1)$ serves as the physical source for the hidden variables $M_i$ in our algorithm. The relevant configurations are the system's occupation-number Fock states $|\mathbf n\rangle=|n_1,n_2,\ldots,n_L\rangle$, with $0\leq n_{\ell}\leq N_{\max}$. Because the effective spectral measure depends on nonlocal configuration mixing, the probability weights $w_i$ assigned to these Fock states are determined self-consistently by the thermodynamic feedback loop.}

\rev{The boundary driving term $F(\hat a_1^\dagger+\hat a_1)$ explicitly breaks the global $U(1)$ particle-number symmetry, injecting and removing bosons and coupling particle-number sectors that are disconnected when $F=0$. In the numerical examples we use $J=1.0$, $F=0.8$, and $U=4.0$: the drive is comparable to the hopping scale while the interaction remains strong. This intermediate competition regime produces substantial hybridization among occupation-number configurations without invoking the trivial atomic limit $J=F=0$ or the extreme interaction-dominated limit.}

\rev{For the parameter regime studied here, direct exact diagonalization of the accessible sizes gives an adjacent-gap ratio approaching the GOE benchmark, while the weighted-spectrum reconstruction exhibits the same GOE-like level-repulsion trend in its surrogate spacing statistics.}


\section{Gaussian Attention-Inspired Mapping for Bosonic Systems}

\rev{Mapping the 1D driven Bose-Hubbard model into a physically constrained feature space motivates a distance-based, attention-inspired relevance map for reconstructing the many-body statistical spectrum.}
\rev{First, consider the deeply non-interacting limit ($U=0$), where hopping can spread amplitudes across the occupation-number basis. Conversely, in the atomic limit ($J=0,F=0$), the Hamiltonian is diagonal in the occupation-number Fock basis, so the Fock states form an eigenbasis independently of the magnitude of $U$. In the intermediate nonintegrable regime, the attention-inspired representation is used to encode configuration relevance associated with the hopping, driving, and interaction scales.}
\rev{For reference, a generic pairwise attention representation models configuration relevance through a normalized attention matrix}
\begin{equation}
{\bf H}	={\rm Attn}({\bf Q},{\bf K},{\bf V})={\bf A}\,{\bf V},
\qquad
{\bf A}=\mathrm{Softmax}(S),
\end{equation}
Here the score matrix \(S\) encodes the relevance between bosonic Fock configurations.
In the simplest compact notation one may write \(S\propto {\bf Q}{\bf K}^T/\sigma_U\), while in the detailed implementation of Appendix~\ref{app:self_attention} we employ a Gaussian distance-based score kernel.
\w{$\sigma_U$ is the standard deviation of the interaction energy distribution, and $\|\mathbf{Q}_i - \mathbf{K}_j\|^2$ quantifies the generalized distance between configuration features in the projected space. 
\rev{The squared-distance score is structurally analogous to the quadratic dependence $(M_i-\mu_t)^2$ used by the thermodynamic feedback in Algorithm~1; the precise reduced global-context realization is stated below and in Appendix~\ref{app:self_attention}.}}
\rev{Here $\mu_t=\sum_{i=1}^{D}w_{i,t}M_i$ is the instantaneous weighted mean interaction energy at iteration $t$.}
 \rev{In the generic Query--Key--Value notation, ${\bf Q} = {\bf X}{\bf W}^Q$, ${\bf K} = {\bf X}{\bf W}^K$, and ${\bf V} = {\bf X}{\bf W}^V$ denote projected configuration features. In the numerical implementation used below, however, these projection matrices are not independently trained: the reduced physical feature is the scalar interaction energy $M_i$, the dynamically pooled context is $\mu_t=\sum_jw_jM_j$, and the optimized variables are the spectral weights $w_i$.}
Here the input data matrix ${\bf X}$ represents the truncated bosonic Fock configurations and its dimension is $(D \times d_{loc})$ where 
$D$ is the total number of Fock states (the effective Hilbert-space dimension), and each row (Fock state) is represented by a feature vector of size $d_{loc}$.
$d_{\mathrm{loc}} = N_{\max} + 1$ denotes the local feature dimension,
i.e., the size of the feature vector used to encode the site-occupation configuration of each state.
 
\rev{The projection matrices $\mathbf W^Q,\mathbf W^K,\mathbf W^V$ have dimensions $(d_{\mathrm{loc}}\times d_k)$, $(d_{\mathrm{loc}}\times d_k)$, and $(d_{\mathrm{loc}}\times d_v)$, respectively, and define the projected feature representation used to construct the Gaussian relevance matrix. In the present coarse-grained implementation these projection matrices are not independently trained through a microscopic wave-function loss. The thermodynamic optimization acts directly on the effective spectral weights $w_i$, while the attention map supplies a structured representation of configuration relevance; details are given in Appendix~\ref{app:self_attention}.}

\rev{The compact score notation in Eq.~(2) is schematic; the implementation used here is the Gaussian distance-based kernel with row-wise softmax normalization defined in Appendix~\ref{app:self_attention}.}
\rev{Operationally, Algorithm~1 uses the reduced global-context score
\[
Q_i^{\rm phys}=M_i,\qquad K_t^{\rm glob}=\mu_t,\qquad
E_i^{(t)}=\eta\frac{\Delta_{\sigma^2}}{\sigma_t^2}(Q_i^{\rm phys}-K_t^{\rm glob})^2,
\]
followed by the normalized exponential update $w_i^{(t+1)}\propto\exp[\ln w_i^{(t)}+E_i^{(t)}]$. Thus, the implemented method retains the distance-based scoring and normalized competition characteristic of attention, but in a pooled physical representation rather than through a separately trained $\mathbf A\mathbf V$ branch. 
}
\rev{Neural quantum-state studies provide a useful point of comparison: FermiNet has been shown to represent both delocalized and localized many-body phases without imposing the phase structure a priori\cite{Cassella2023}. Our construction is deliberately more restricted: it does not variationally optimize a microscopic many-body wave function, but uses an attention-inspired feature map to organize a coarse-grained spectral measure over Fock configurations.}

\w{The scaling factor $\sigma_U$ effectively normalizes the macroscopic energy differences in the attention scores. Crucially, the coherent boundary driving term $F(\hat{a}_1^\dagger + \hat{a}_1)$ explicitly breaks the global $U(1)$ particle-number symmetry. By removing the block-diagonal restrictions of fixed particle-number sectors, this driving field opens extensive resonance channels, forcing the system to scatter into previously inaccessible Hilbert space regions and intrinsically generating many-body ergodicity.}

\rev{To illustrate the transition from integrability to chaos, we analyze the interplay between the diagonal interaction term and the off-diagonal hopping and driving operators. The target second moment is evaluated from $\mathrm{Tr}(H)$ and $\mathrm{Tr}(H^2)$, whereas the full eigenvalues and eigenvectors are computed only for the accessible sizes used as independent ED benchmarks. The finite drive and hopping mix occupation-number configurations and connect particle-number sectors in the truncated Hilbert space.}


\section{Algorithm and Renormalization Group Flow}

Calculating the full spectrum for large lattice size $L$ and high particle truncation $N_{max}$ is computationally prohibitive.
Therefore, we employ an algorithm based on step length modulation to predict the spectral weights.
The core of our method relies on a renormalization group (RG) flow analysis of the statistical variance.
\rev{More explicitly, for a dense $D\times D$ matrix a complete eigendecomposition has the standard cubic cost $\mathcal O(D^3)$ and quadratic storage $\mathcal O(D^2)$. By contrast, the second spectral moment required by the reconstruction can be obtained without diagonalization from
\begin{equation*}
\mathrm{Tr}(H^2)=\sum_{i,j}H_{ij}H_{ji}=\sum_{i,j}|H_{ij}|^2,
\end{equation*}
where the last equality holds for the Hermitian Hamiltonian considered here. Hence, once $H$ is available in a standard sparse representation, the trace evaluation requires $\mathcal O(\mathrm{nnz}(H))$ operations. For the nearest-neighbor Bose--Hubbard chain at fixed local cutoff $N_{\max}$, each Fock configuration is connected to only $\mathcal O(L)$ other configurations, so $\mathrm{nnz}(H)=\mathcal O(LD)$. Once the hidden energies $M_i$ are constructed, each Algorithm~1 update of the weighted mean, variance, modulation exponent, and normalization is $\mathcal O(D)$, giving a reconstruction cost of approximately $\mathcal O[\mathrm{nnz}(H)+T_{\max}D]$ and sparse storage $\mathcal O[\mathrm{nnz}(H)+D]$. Thus the gain is a reduction in the polynomial dependence on the Hilbert-space dimension relative to dense full-spectrum diagonalization; it does not change the exponential dependence of $D$ on $L$.}
\rev{We introduce a dimensionless scaling parameter $\chi$ as an auxiliary scale for the accessible reconstructed state space. It is not identified with a microscopic coupling such as $J^{-1}$; increasing $\chi$ corresponds to increasing participation in the reconstructed measure.}

We define a dimensionless scaling function $f_{\chi}$, which characterizes the partition of spectral weight between coupled states.
\rev{Based on the invariance of the physical variance $\sigma^2$ under the reparameterization of weights, we use the flow relation $\sigma^{2}=f_\chi/(1-\partial_\chi f_\chi)$, whose derivation is summarized in Appendix~\ref{app:rg}.}
\rev{Considering a binary group of weights $\{w_1,w_2\}$ associated with two Fock configurations whose diagonal interaction energies are set by $U$ and which may be mixed by the off-diagonal hopping and driving terms $J$ and $F$, the flow towards the fixed point implies:}
\begin{equation}
	\begin{aligned}
		\label{limf}
		\lim_{\chi\rightarrow\infty}f_{\chi} = \frac{1}{2}, \quad \text{and} \quad
		\frac{1}{2} = f_{\chi} - f_{\chi}^{2} \partial_{\chi} \left( \frac{1}{2f_{\chi}} \right).
	\end{aligned}
\end{equation}
\rev{The fixed-point value $1/2$ corresponds to maximal mixing of the illustrative two-state weight pair. This auxiliary large-$\chi$ limit should not be identified with the physical atomic limit $U/J\to\infty$: the latter is Fock-space localized because the on-site interaction is diagonal in the occupation basis. Here $\chi$ only parametrizes the participation scale of the reconstructed probability measure.}

\subsection{Criterion and $\beta$-Ensemble}

We utilize the Dyson index $\beta_{en}$ to characterize the statistical phase transition.
For the driven Bose-Hubbard model in Eq. (\ref{Hb}), while real parameters imply time-reversal symmetry and GOE statistics ($\beta_{en}=1$), the interplay between the driving field $F$ and interaction $U$ can effectively break integrability.
Our algorithm lifts the ensemble index from the localized (Poisson, $\beta_{en} \to 0$) limit to the chaotic limit.
\w{To quantify the emergence of level repulsion, we compare the statistics of our
reconstructed effective spectrum with the random-matrix surmise.
In Dyson's $\beta$-ensemble, the joint probability density of $\rev{D}$ ordered levels
$\{\lambda_i\}$ is\cite{Tracy C A,Dumitriu I,Mehta}
\begin{equation}
P_{\beta}(\{\lambda\}) =
C_{\beta,\rev{D}}\prod_{i<j}|\lambda_i-\lambda_j|^{\beta_{\mathrm{en}}}
\exp\!\left[-\frac{1}{2f}\sum_{i=1}^{\rev{D}}(\lambda_i-\bar{\lambda})^2\right],
\label{eq:JPDF_beta}
\end{equation}
where $f$ is the global spectral variance and
$\bar{\lambda}=\frac{1}{\rev{D}}\sum_i \lambda_i$.
In our framework we do not compute the exact eigenvalues directly; instead we
construct a weighted discrete spectral measure
$\rho_{\mathrm{eff}}(E)=\sum_i w_i\,\delta(E-M_i)$ whose first and second moments are constrained to match the target many-body spectrum and here \rev{$w_i$ is the coarse-grained probability weight over the diagonal interaction states.}
The effective repulsion parameter $\beta_{\mathrm{en}}$ is then extracted by fitting
the unfolded spacing statistics derived from $\rho_{\mathrm{eff}}$ to the $\beta$-ensemble
surmise, and we observe $\beta_{\mathrm{en}}$ evolves from $0$ to $1$ under the
iterative learning dynamics, indicating the onset of Wigner--Dyson-like correlations.
As a phenomenological reference, we model the correlations of the reconstructed
effective levels $\{M_i\}$ using a Dyson $\beta$-ensemble form
\begin{equation}
P_{\beta}(\{M\}) =
C_{\beta,\rev{D}}\prod_{i<j}|M_i-M_j|^{\beta_{\mathrm{en}}}
\exp\!\left[-\frac{1}{2f}\sum_{i=1}^{\rev{D}}(M_i-\mu)^2\right],
\label{eq:JPDF_beta_M}
\end{equation}
where $\mu$ is the (weighted) spectral mean and $f$ is the corresponding
raw weighted variance (standard second central moment). With normalized weights $\sum_i w_i=1$, we use
\begin{equation}
f=\sum_{i=1}^{D}w_i(M_i-\mu)^2=\sum_{i=1}^{D}w_i M_i^2-\mu^2,
\label{eq:weighted_mean_var}
\end{equation}
 \rev{which represents the raw variance of the spectral distribution before correcting for the effective sample size. For the $i$-th Fock configuration, $M_i=\sum_{\ell=1}^{L}\frac{U}{2}n_{i,\ell}(n_{i,\ell}-1)$ is its bare diagonal interaction energy, and $\mu=\sum_iw_iM_i$ is the weighted mean.}
After unfolding the reconstructed measure, the effective repulsion parameter $\beta_{\mathrm{en}}$
is inferred by fitting the resulting spacing statistics to the $\beta$-ensemble surmise.
}

\rev{The full-Hamiltonian variance $\sigma_{\mathrm{target}}^2$ is governed by both the diagonal interaction term and the off-diagonal hopping and driving terms. It is evaluated from the trace invariants of the full Hamiltonian rather than from its diagonalized spectrum. To construct an operational target compatible with the diagonal hidden variables $M_i$, we subtract the approximate kinetic broadening, as detailed in Sec.~\ref{feedback}, and use $\sigma_{\mathrm{target,eff}}^2\simeq\sigma_{\mathrm{target}}^2-\mathcal O(LJ^2)$.}


\subsection{Method: Variance Stability in Fock Space}

Here we demonstrate the principle of our algorithm for predicting the sample weight distribution (spectral density) of the driven Bose-Hubbard model through the mean value and variance estimated in a limited Hilbert space subspace.
During the steps of automatic iterations, the criterion is to match the mean value and variance of the predicted spectrum with the reduced one (irreducible group of Fock states, which has a much smaller sample size).

\rev{For the binary illustration below we consider two Fock configurations $|n\rangle$ and $|m\rangle$ with diagonal interaction energies determined by $U$; any physical mixing between distinct configurations is generated by the off-diagonal hopping and driving terms $J$ and $F$. The variance identity itself concerns only the redistribution of their statistical weights.}

For example, for a group $\{M_{1},M_{2}\}$ (representing eigenenergies of the doublet) weighted by $\{w_{1},w_{2}\}$, the variance reads (using the bias-corrected sample weighted variance)
\begin{equation}
	\begin{aligned}
		\label{unbi}
		\sigma^{2}
		=\frac{\sum_{i}w_{i}(M_{i}-\mu)^2}{\sum_{i}w_{i}-\frac{\sum_{i}w_{i}^{2}}{\sum_{i}w_{i}}},
	\end{aligned}
\end{equation}
where $\mu=\sum_{i}M_{i}w_{i}/\sum_{i}w_{i}$ is the weighted mean energy.
\rev{For normalized weights $\sum_iw_i=1$, the same estimator is written consistently in the notation used throughout the manuscript as $\sigma^{2}=\frac{\sum_iw_i(M_i-\mu)^2}{1-\sum_iw_i^2}$.}
Next we consider only the case $\sum_{i}w_{i}=1$.
Then we introduce another configuration with modified weight distribution $\{w'_{1},w'_{2}\}:=\{w_{1}+w_{d},w_{2}-w_{d}\}$, representing a redistribution of probability amplitude driven by the external field $F$.
These two groups have the same variance
\begin{equation}
	\begin{aligned}
		\sigma^{2}&=\frac{1}{2}(\rev{M_{1}}-\rev{M_{2}})^{2} \\
		&=\frac{(\rev{M_{1}}-\rev{M_{2}})^{2}\prod_{i}w_{i}}{1-\sum_{i}w_{i}^2} \\
		&=\frac{(\rev{M_{1}}-\rev{M_{2}})^{2}(\prod_{i}w_{i}-w_{d}^{2}+w_{d}(1-2w_{1}))}
		{1-\sum_{i}w'_{i}{}^2}.
	\end{aligned}
\end{equation}
This stability can be explained by expressing this common variance $\sigma^{2}$ in terms of a limiting result of the infinitely scaled variable $\chi$ (which here corresponds to the scaling of the effective Hilbert space dimension), having the form $\sigma^{2}:=\lim_{\chi\rightarrow \infty}f =\frac{f}{1-\partial_{\chi}f}$,
\rev{Here $\partial_\chi f$ is identified with the inverse participation ratio (IPR) of the reconstructed weights, $\mathrm{IPR}_w\equiv\sum_{i=1}^{D}w_i^2$. In the delocalized regime $\partial_\chi f\sim1/D$, so the correction factor tends to unity and $\sigma^2\simeq f$. In the localized limit $\partial_\chi f=\sum_iw_i^2\to1$, so the factor $(1-\partial_\chi f)^{-1}$ becomes singular, signaling the breakdown of an effective many-state ensemble description. The associated participation number $N_{\mathrm{eff}}\equiv1/\mathrm{IPR}_w$ therefore varies from approximately $1$ in a localized distribution to approximately $D$ in a maximally spread distribution. Since $S_2=-\ln(\mathrm{IPR}_w)$, the delocalized scaling $\mathrm{IPR}_w\sim D^{-1}$ corresponds to $S_2\sim\ln D$.}
\rev{For the present reconstruction, no additional logistic, Landau-parameter, or geometric singularity interpretation is required. We use $\mathrm{IPR}_w=\sum_iw_i^2$ only as a participation diagnostic and $1-\mathrm{IPR}_w$ as the finite-weight correction factor entering the bias-corrected variance. Thus $w_k\to1$ corresponds to the single-state localized limit, whereas $w_i\simeq1/D$ corresponds to a broadly distributed measure.}

\rev{In the maximally spread reconstructed limit, $w_i\simeq1/D$ and $\sum_iw_i^2\simeq1/D$, so the bias correction becomes negligible and the corrected variance approaches the raw weighted second moment. Increasing $\chi$ is therefore associated with broader participation of the reconstructed measure and a smaller $\partial_\chi f$.}
\rev{Here $\chi$ is used only as an effective scale variable for the accessible reconstructed state space; for fixed local dimension, its large-space behavior may be associated with a logarithmic volume such as $\chi\sim\ln D$. We do not identify $\chi$ with a field-theoretic ultraviolet cutoff.}
\rev{Accordingly, in the maximally delocalized limit $\partial_\chi f\sim\sum_i(1/D)^2=1/D$, which vanishes as the accessible Hilbert-space dimension grows.}
\rev{The relation $S_2=-\ln(\mathrm{IPR}_w)$ provides a convenient interpretation of the large-$D$ limit: as the reconstructed measure spreads over more hidden-variable states, $S_2$ increases and the finite-sample correction becomes negligible. This entropy interpretation is used only as a diagnostic of participation in the reconstructed measure and not as an additional microscopic field-theory assumption.}

In the above expression, the evolution of the variance can be analyzed by treating the weights as dynamic variables under a renormalization group (RG) flow parameterized by the scale $\chi$.
For a binary group satisfying normalization $\sum w_i = 1$, the sum of squared weights relates to their product $w_1 w_2$ via
\begin{equation}
	\begin{aligned}
		\sum_{i}w_{i}^2 = 1 - 2w_{1}(1-w_{1}).
	\end{aligned}
\end{equation}
We define the scaling function $f_{\chi} = 2w_{1}(1-w_{1})$ (scaled to have a maximum of $1/2$ at the symmetric fixed point). The fixed point of the flow is determined by the condition of maximal entropy (or maximal variance mixing), implying
\begin{equation}
	\begin{aligned}
		\label{o0}
		&\lim_{\chi\rightarrow\infty}f_{\chi}=\frac{1}{2},\\
		&\partial_{\chi}f_{\chi} = 1 - 2f_{\chi}.
	\end{aligned}
\end{equation}
The second line indicates that the flow of $f_\chi$ is driven by its deviation from the fixed point value $1/2$, characteristic of a linear RG flow near a critical point.

\rev{Regarding the illustrative two-state gap $\Delta_M^2\equiv(M_1-M_2)^2$, we consider a representation where the effective energy units become scale-dependent as the accessible Fock-space measure is varied.} As the system flows toward the chaotic fixed point, the dynamical scaling function $f_\chi$ obeys a generic linear response relation. Since the RG flow must stall precisely at the maximum entropy state (the fixed point $f_\chi \to 1/2$), its scale derivative to leading order is given by $\partial_\chi f_\chi = 1 - 2f_\chi$.

Due to the strong hybridization induced by the driving field $F$ and the interaction $U$, the effective energy gaps are dynamically renormalized as the Hilbert space is explored. 
\rev{The finite local-occupation cutoff $N_{\max}$ truncates the local Hilbert space and regularizes this finite-dimensional reconstruction, enforcing the saturation of the scaling function}
\begin{equation}
    \lim_{\chi \to \infty} \left| \frac{1}{2} - \frac{f_\chi}{1 - \partial_\chi f_\chi} \right| = 0.
\end{equation}
Substituting the dynamical flow equation into the energy gap derivative, we obtain
\begin{equation}
	\begin{aligned}
		\label{uu}
		\partial_{\chi}\rev{\Delta_M^2}
		=\frac{1-2f_{\chi}}{f_{\chi}}[1-\rev{\Delta_M^2}].
	\end{aligned}
\end{equation}
Using the established fixed-point response relation $1-2f_\chi = \partial_\chi f_\chi$, Eq.(\ref{uu}) simplifies to the logarithmic derivative form
\begin{equation}
	\begin{aligned}
		\label{pp98}
		\frac{1-\rev{\Delta_M^2}}{\partial_{\chi}\rev{\Delta_M^2}}
		= \frac{f_{\chi}}{\partial_{\chi}f_{\chi}}.
	\end{aligned}
\end{equation}
This differential equation reveals that the energy gap squares $\rev{\Delta_M^2}$ and the weight distribution function $f_\chi$ share the same scaling exponents near the chaotic fixed point.
Consequently, in the limit $\chi \to \infty$, the normalized gap $\rev{\Delta_M^2}$ flows towards unity (in dimensionless units), while the statistical variance stabilizes.
This explains why the variance of the Fock states remains invariant under the iterative modification of weights: the renormalization of the interaction strength $U$ is exactly compensated by the scaling of the accessible Hilbert space measure.

\section{Thermodynamic Feedback Control of Spectral Variance}

\rev{Here we demonstrate the principle of our algorithm for reconstructing the spectral-weight distribution of the driven Bose--Hubbard model. Rather than repeatedly diagonalizing the full Hamiltonian, the reconstruction iteratively adjusts the probability weights $w_i$ assigned to the diagonal interaction energies $M_i$ to match the prescribed statistical moments.}

The core of the method is a negative feedback control loop inspired by non-equilibrium thermodynamics. We treat the spectral weight distribution for all $D$ Fock states (or hidden variable states) $\mathbf{W} = \{w_i\}= [w_1, w_2, \dots, w_D]^T$ as a dynamic ensemble evolving under a fictitious potential,
where $w_i$ corresponds to the probability weight of the $i$-th state.
The objective is to drive the current weighted variance $\sigma_t^2$
\rev{towards the operational effective target variance $\sigma_{\mathrm{target,eff}}^2$, while the full-Hamiltonian second moment $\sigma_{\mathrm{target}}^2$ is evaluated from trace invariants.}

\subsection{Adaptive Modulation Potential}

To control the second moment (variance) of the distribution, we introduce a modulation potential $V(M_i)$ that acts on the weights. The update rule at iteration $t$ is given by the Boltzmann-like factor
\begin{equation}
	w_i^{(t+1)} = \frac{w_i^{(t)} e^{E_i^{(t)}}}{Z_t},
    \label{14}
\end{equation}
where $Z_t=\sum_j w_j^{(t)} e^{E_j^{(t)}}$ is the partition function (normalization) and $E_i^{(t)}$ is the feedback energy.
Since the variance measures the spread of the distribution relative to its mean $\mu_t$, the conjugate field required to manipulate the variance must be quadratic in the energy deviation $(M_i - \mu_t)$.

\rev{We define the feedback energy $E_i$ proportional to the operational variance mismatch $\Delta_{\sigma^2}=\sigma_{\mathrm{target,eff}}^2-\sigma_t^2$}
\begin{equation}
	\label{eq:feedback}
	E_i^{(t)} = \eta  \frac{\Delta_{\sigma^2}}{\sigma_t^2}  (M_i - \mu_t)^2,
\end{equation}
where $\eta$ is a learning rate controlling the stiffness of the feedback.
When the predicted distribution is too narrow ($\Delta_{\sigma^2} > 0$), the prefactor in Eq.~(\ref{eq:feedback}) is positive. This generates an inverted quadratic (deconfining) potential 
$e^{+|E_i|}$ that amplifies the weights at the edges of the spectrum (where $|M_i - \mu_t|$ is large). \rev{This effectively flattens (heats) the distribution, pushing probability mass outward to broaden the envelope and increase the variance.}
 Conversely, when the distribution is too wide ($\Delta_{\sigma^2} < 0$), the prefactor becomes negative. This applies a confining quadratic (harmonic) potential $e^{-|E_i|}$ that suppresses the tails and concentrates the weights near the center, thereby cooling the distribution and reducing the variance.
\rev{The two sign-dependent branches follow directly from Eq.~(\ref{eq:feedback}): the positive branch broadens the distribution, whereas the negative branch compresses it toward the current weighted mean energy.}

\rev{We then define the loss function as the squared difference between the effective trace-based target used by the reconstruction and the current predicted variance}
\begin{equation}
	\mathcal{L}(\mathbf{w}) = \frac{1}{2} \left( \rev{\sigma_{\mathrm{target,eff}}^2} - \sigma_t^2(\mathbf{w}) \right)^2,
    \label{loss}
\end{equation}
where $\sigma_{t}^2(\mathbf w)$ is the bias-corrected predicted variance (standard second central moment) at iteration $t$, defined explicitly in Eq.\ref{target}.
The dynamically predicted variance used in the feedback loop is then taken as the bias-corrected estimator.

\w{The objective of the thermodynamic feedback is to minimize the discrepancy between the macroscopic spectral envelope of the interacting system and the predicted distribution.
As discussed in the context of the replica method and effective sample sizes, the raw variance estimator suffers from bias in strongly interacting regimes. To enhance numerical stability near the chaotic transition and properly account for the inverse participation ratio constraints, we apply the dynamically bias-corrected predicted variance which serves as the actual operational metric in Algorithm 1. 
Taking the partial derivative of $\mathcal{L}(w)$ with respect to the weight parameter yields the gradient that drives the exponential update. The resulting local feedback potential $E_i$ applied to each hidden variable $M_i$
 and $\Delta_{\sigma^2}$ determines the direction of the optimization. Crucially, the presence of $\sigma_{t}^2$ in the denominator normalizes the gradient, ensuring that the feedback force is scaled relative to the current distribution width, thereby preventing numerical divergence when comparing broad versus narrow ensembles.}

The algorithm minimizes the loss $\mathcal{L}(\mathbf{w})$ and searching for the optimal configuration of the vector $\mathbf{w}$ (blue points in Fig.1 and Fig.2) that makes the current variance match the target variance.
To reconstruct the Wigner-Dyson statistics, the system must evolve to minimize this loss $\mathcal{L}$.
The evolution of the weights follows a gradient descent flow in the logarithmic domain (to ensure positivity of probabilities, similar to the exponentiated gradient algorithm). The update force $F_i$ acting on the $i$-th state is proportional to the negative gradient of the loss with respect to the log-weight
\begin{equation}
\begin{aligned}
&	F_i = - \eta \frac{\partial \mathcal{L}}{\partial \ln w_i}
    = - \eta w_i \frac{\partial \mathcal{L}}{\partial w_i}
    = \eta (\rev{\sigma_{\mathrm{target,eff}}^2}-\sigma_t^2) \frac{\partial \sigma_t^2}{\partial w_i},\\
\frac{\partial \mathcal{L}}{\partial w_i} 
&=\frac{\partial \mathcal{L}}{\partial (-\sigma_t^2)}
\frac{\partial (-\sigma_t^2)}{\partial w_i} \\
&= w_i(\rev{\sigma_{\mathrm{target,eff}}^2}-\sigma_t^2) \cdot \frac{\partial (-\sigma_t^2)}{\partial w_i} \\
&= - w_i(\rev{\sigma_{\mathrm{target,eff}}^2}-\sigma_t^2) \cdot \frac{\partial}{\partial w_i} \left[ \sum_k w_k (M_k - \mu)^2 \right]\\
&\approx - w_i\Delta_{\sigma^2} \cdot (M_i - \mu)^2
\end{aligned}
\end{equation}
where $F_i$ is treated as natural gradient (Riemannian gradient)
\w{
$\frac{\partial \sigma_t^2}{\partial w_i} = (M_i - \mu_t)^2 - \mu_t^2$ implies that the sensitivity of the variance to a specific weight $w_i$ is dominated by its squared distance from the mean.}
\rev{Hence, the farther a state lies from the current mean, the more strongly its weight affects the variance.}
While for the evolution of weights follows the exponentiated gradient descent (EGD) scheme
(with update rule $\ln w_{new} = \ln w_{old} - \eta \frac{\partial \mathcal{L}}{\partial w_{old}}$ and $w_{new} = w_{old} \cdot \exp\left( -\eta \frac{\partial \mathcal{L}}{\partial w_{old}} \right)$),
which is tailored for optimization over a probability simplex, the update force $F_i$ in the exponent corresponds to the negative Euclidean gradient of the loss
\begin{equation}
F_i = - \eta \frac{\partial \mathcal{L}}{\partial w_i} = \eta \Delta_{\sigma^2} (M_i - \mu_t)^2,
\label{18}
\end{equation}
\rev{which is the thermodynamic modulation potential $E_i$ used in Algorithm~1. The update approaches a stationary variance-matching condition as $\Delta_{\sigma^2}\to0$ and $F_i\to0$, so that}
$\frac{dw_i}{dt} = w_i \cdot F_i\to 0$.
\rev{When $\Delta_{\sigma^2}>0$, states farther from the current mean receive the larger positive modulation and the variance increases; when $\Delta_{\sigma^2}<0$, the tails are suppressed and the variance decreases.}

In the deep learning framework of a standard Transformer, to update the query matrix $W^Q$, the chain rule is used to propagate the gradients from the final Loss $\mathcal{L}$ backwards step by step
$\frac{\partial \mathcal{L}}{\partial W^Q} = \frac{\partial \mathcal{L}}{\partial \mathbf{w}} \cdot \frac{\partial \mathbf{w}}{\partial \text{Score}} \cdot 
\frac{\partial \text{Score}}{\partial Q} \cdot
\frac{\partial Q}{\partial W^Q}$, 
with the attention weights $\mathbf{w}$ corresponding to the $w_i$. 
$\frac{\partial \mathcal{L}}{\partial \mathbf{w}}$ is the gradient with respect to the attention weights, while the subsequent term $\frac{\partial \mathbf{w}}{\partial \text{Score}}$ represents the derivative of the Softmax function.

The relationship between our thermodynamic update rule and the standard backpropagation algorithm can be understood via the chain rule.
In a standard Transformer, the gradient descent update for the query matrix is given by $W^Q \leftarrow W^Q - \eta \frac{\partial \mathcal{L}}{\partial W^Q}$ (according to the update rule of projected gradient descent (PGD)).
By expanding the gradient term $\frac{\partial \mathcal{L}}{\partial W^Q} = \frac{\partial \mathcal{L}}{\partial \mathbf{w}} \frac{\partial \mathbf{w}}{\partial W^Q}$, we identify that our feedback force $F_i$ corresponds precisely to the first term in this chain $F_i \propto -\frac{\partial \mathcal{L}}{\partial w_i}$.
Therefore, instead of updating the query weight matrix $W^Q$ through the attention mechanism
(error signal flow backwards),
our algorithm effectively performs variational inference directly in the probability space of the attention weights, bypassing the computational overhead of training the underlying projection matrices while preserving the physical objective of moment matching.


\w{In the deeply chaotic regime, where the state is extensively delocalized across the Fock space, the inverse participation ratio is vanishingly small ($\sum_{i=1}^D w_{i,t}^2 \ll 1$). Under this limit, the dynamically predicted variance $\sigma_{t}^2$ naturally reduces to the raw second central moment $f_t = \sum_{i=1}^D w_{i,t} (M_i - \mu_t)^2$. However, to ensure general numerical stability across the entire parameter space---especially near the localization transition where $\sum_{i=1}^D w_{i,t}^2$ becomes macroscopic---Algorithm 1 strictly employs the full bias-corrected estimator $\sigma_{t}^2$. To successfully dynamically reconstruct the macroscopic envelope required for Wigner-Dyson statistics, the feedback mechanism must evolve the weights to minimize the loss $\mathcal{L}$, forcing $\sigma_t^2$ to converge toward $\sigma_{\mathrm{target,eff}}^2$.}

\subsection{Target Variance Adjustment}
\label{feedback}

\rev{It is crucial to distinguish the full-Hamiltonian target second moment from the ED validation data. For a given set of parameters $(U,J,F,L,N_{\max})$, the full spectral variance is evaluated directly from trace invariants, without computing the eigenspectrum,}
\begin{equation}
\sigma_{\mathrm{target}}^2 = \frac{1}{D}\mathrm{Tr}(H^2)-\left[\frac{1}{D}\mathrm{Tr}(H)\right]^2.
\label{DH}
\end{equation}
Here $\sigma_{\mathrm{target}}^2$ is identical to the exact eigenvalue variance $\frac{1}{D}\sum_{n=1}^D E_n^2 - (\frac{1}{D}\sum_{n=1}^D E_n)^2$, yet it is evaluated directly via trace invariants without computing the eigenvalues themselves.
\rev{This quantity includes the diagonal interaction contribution and the off-diagonal hopping and driving contributions. However, }
because the hidden variables $M_i$ are constructed only from the diagonal interaction energies, the off-diagonal hopping and driving contribute an additional broadening to the true spectrum that cannot be represented exactly within diagonal hidden-variable basis. 

To prevent the algorithm from forcing the weights to the extreme boundaries \rev{(boundary-dominated weight accumulation observed when the diagonal basis is forced to reproduce the full spectral width)} in an attempt to match the full width, we introduce an effective target variance 
\begin{equation}
\sigma_{\mathrm{target,eff}}^2 \approx \sigma_{\mathrm{target}}^2 - \sigma_{\mathrm{kin}}^2,
\end{equation}
which empirically relaxes the target width to match the diagonal-basis-compatible spectral envelope.
\rev{In the numerical implementation the kinetic broadening is approximated by $\sigma_{\mathrm{kin}}^2\simeq2LJ^2$, so that the feedback uses $\sigma_{\mathrm{target,eff}}^2\simeq\sigma_{\mathrm{target}}^2-2LJ^2$. This operational projection prevents the diagonal hidden-variable weights from being forced artificially to the spectral boundaries.}

\rev{The weighted-spectrum reconstruction is used as a phenomenological surrogate for the GOE-like level-repulsion trend. Appendix~B provides a complementary random-matrix connection between spectral repulsion and eigenvector structure, without identifying the surrogate weights with microscopic eigenvector intensities.}

\section{algorithm and simulation result}

\begin{algorithm}[H]
	\caption{\rev{Attention-Inspired Spectral Prediction via Thermodynamic Feedback}}
	\label{alg:spectral_prediction}
	\DontPrintSemicolon
	\KwIn{Hamiltonian ($U, J, F, L$), Basis $\mathbf{X}$, Tolerance $\epsilon$, Rate $\eta$.}
	\KwOut{Spectral weights $\mathbf{W}$, Variance $\sigma_{\Delta}^2$.}
	
	\textbf{Phase 1: Initialization} \\
\rev{Construct the physical hidden-variable map: ${\bf M} \leftarrow \operatorname{diag}(\mathbf{H}_{int})$}\;
	\rev{Evaluate $\sigma^2_{\mathrm{target}}$ from $\mathrm{Tr}(H)$ and $\mathrm{Tr}(H^2)$ and construct $\sigma^2_{\mathrm{target,eff}}$}\;
	Initialize weights: $w_i \leftarrow 1/D$, $\lambda_i \leftarrow 0$\;
	
	\BlankLine
	\textbf{Phase 2: Feedback Optimization Loop} \\
	\For{$t = 1$ \KwTo $T_{max}$}{
		Calculate current weighted mean $\mu_t$ and variance $\sigma_t^2$: \\
		$\mu_t = \sum w_i M_i, \quad \sigma_t^2 = \frac{\sum w_i (M_i - \mu_t)^2}{1 - \sum w_i^2}$\;
		
		\Comment{Check Convergence}
		\If{$|\sigma_t^2-\rev{\sigma_{\mathrm{target,eff}}^2}|<\epsilon$}{
			\textbf{break}\;
		}
		
		\Comment{Calculate Feedback Force}
		Compute variance mismatch error: $\Delta_{\sigma^2}=\rev{\sigma_{\mathrm{target,eff}}^2}-\sigma_t^2$\;
		
		\Comment{Adaptive Modulation (Heating/Cooling)}
		Construct modulation exponent $E_i$ for each state $i$: \\
		$E_i \leftarrow \eta \cdot \frac{\Delta_{\sigma^2}}{\sigma_t^2} \cdot (M_i - \mu_t)^2$\;
		
		\eIf{$\Delta_{\sigma^2} > 0$}{
			\Comment*[r]{Heat: Inverted Gaussian to expand width}
			(Implicitly $E_i > 0$ enhances tails)
		}{
			\Comment*[r]{Cool: Normal Gaussian to shrink width}
			(Implicitly $E_i < 0$ suppresses tails)
		}
		
		\Comment{Update Weights (Softmax ensures normalization)}
		$\mathbf{W} \leftarrow \text{Softmax}(\ln(\mathbf{W}) + \mathbf{E})$\;
	}
	
	\BlankLine
	\textbf{Phase 3: Result Extraction} \\
	Compute effective ensemble index $\beta_{en}$ from final distribution\;
	\Return $\mathbf{W}, \beta_{en}, \sigma_t^2$\;
\end{algorithm}

The Algorithm 1 employs a negative feedback control loop. It calculates the variance mismatch $\Delta_{\sigma^2}$ at every step.
It introduces a sign-dependent modulation potential: 
If the current variance is too small ($\Delta_{\sigma^2} > 0$ with narrow distribution), it applies an inverted Gaussian potential ($\exp[+x^2]$), 
enhances the weights farther from the current weighted mean and therefore
broadens the distribution,
corresponding to the Heating process. 
Here $x$ represents the deviation of the interaction energy from the current mean $x_i \sim (M_i - \mu_t)$, with $M_i$ the interaction energy of the $i$-th Fock state (the fixed horizontal position of the blue point) and $\mu_t=\sum_{i=1}^{D} w_{i,t} M_i$ the current weighted mean energy (center of mass of the blue distribution at iteration $t$). 
This pushes weights toward the edges, increasing the variance.
If the current variance is too large ($\Delta_{\sigma^2} < 0$ with wide distribution), it applies a normal Gaussian potential ($\exp[-x^2]$,
a negative quadratic potential) and
trap the weights toward the center, decreasing the variance, corresponding to the cooling process.
This error-based feedback control
applies the adaptive function ($\exp[\pm x^2]$) as the modulation function,
\rev{which stabilizes the macroscopic variance target; the local GOE-like surrogate spacing statistics are assessed separately after the stochastic roughening of Algorithm~2.}

\rev{The algorithm utilizes a thermodynamic feedback potential $E_i = \eta \cdot \frac{\Delta_{\sigma^2}}{\sigma_{t}^2} \cdot (M_i - \mu_t)^2$, where the sign is determined by the variance mismatch. This imposes a quadratic confining (or deconfining) potential in probability space and drives the reconstructed measure toward the effective target variance (Wigner-Dyson ensemble statistics).}
$E_i$ is the modulation potential which
 updates the the weight of state $i$ as $w_{new} = w_{old} \times e^{E_i}$.
$\eta$ is the learning rate controlling the thermodynamic stiffness of the feedback: High $\eta$ corresponds to fast reaction, but might oscillate, while low $\eta$ corresponds to slow and smooth convergence (overdamped).
\rev{$\Delta_{\sigma^2}=\sigma_{\mathrm{target,eff}}^2-\sigma_t^2$ is the operational error signal used by Algorithm~1.}
\begin{equation}
\sigma^2_{t} = \frac{\sum_{i=1}^{D} w_{i,t}(M_i - \mu_{t})^2}{1 - \sum_{i=1}^{D} w_{i,t}^2}
\label{target}
\end{equation}
is the dynamically (bias-corrected) predicted variance at iteration step $t$.
In the feedback potential, it serves as a scale-setting denominator, so that the update remains dimensionless and numerically stable relative to the current width of the distribution.
The term $(M_i - \mu_t)^2$ ensures that the heating/cooling force is strongest at the edges (where $M_i$ is far from $\mu_t$) and zero at the center (where $M_i \approx \mu_t$).
The denominator $1-\sum_i w_{i,t}^2$ is the reliability-weight correction factor, while the use of $\sigma_{t}^2$ in the feedback potential makes the update dimensionless and ensures that the feedback strength is scaled relative to the current width of the distribution.

\w{To clarify the role of Algorithm~1, we decompose the Hamiltonian into its diagonal and off-diagonal parts,
\begin{equation}
H = M + K,
\end{equation}
where $M$ contains the diagonal interaction energies in the Fock basis, and $K$ contains the off-diagonal hopping and driving terms. The exact spectral variance of the full Hamiltonian is defined in Eq.\ref{DH}.
Since $M$ is diagonal and $K$ is purely off-diagonal in this basis, one has
\begin{equation}
\mathrm{Tr}(K)=0,
\qquad
\mathrm{Tr}(MK)=\mathrm{Tr}(KM)=0.
\end{equation}
Therefore,
\begin{equation}
\mathrm{Tr}(H^2)=\mathrm{Tr}(M^2)+\mathrm{Tr}(K^2),
\end{equation}
and hence
\begin{equation}
\sigma_{\mathrm{target}}^2
=
\left[
\frac{1}{D}\mathrm{Tr}(M^2)
-
\left(\frac{1}{D}\mathrm{Tr}(M)\right)^2
\right]
+
\frac{1}{D}\mathrm{Tr}(K^2).
\label{eq:variance_decomposition}
\end{equation}
\rev{Equation~(\ref{eq:variance_decomposition}) shows that the off-diagonal couplings contribute a finite broadening to the full second spectral moment. Algorithm~1 does not resolve the microscopic interference generated by $K$. Because its hidden variables contain only the diagonal interaction energies, the feedback instead matches the projected target $\sigma_{\mathrm{target,eff}}^2$ defined in Sec.~\ref{feedback}. In this sense the deterministic reconstruction is a coarse-grained surrogate for the diagonal-basis-compatible spectral envelope; the microscopic local fluctuations needed for the surrogate spacing statistics are supplied separately by Algorithm~2.}}

	\begin{figure}
		\centering
\includegraphics[width=0.9\linewidth]{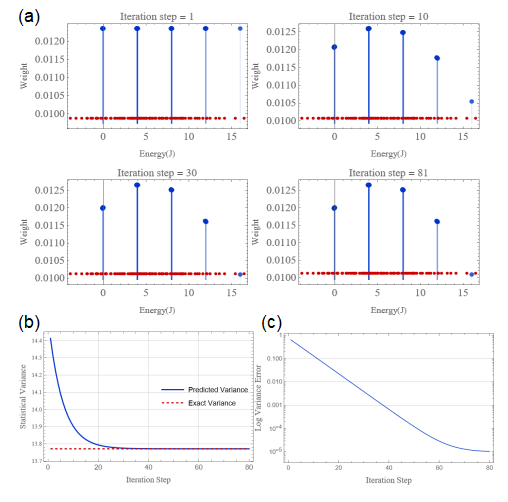}
		\caption{
Simulation results of Algorithm~1 for the driven Bose-Hubbard chain in the overdamped regime.  Parameters are $L=4$, $N_{\max}=2$, $J=1.0$, $U=4.0$, $F=0.8$, learning rate $\eta=0.1$, and the total number of iterations is $80$. 
(a) Evolution of the dynamically learned weights distributed over the hidden interaction energies $M_i$.
Notably, the highest interaction energies consistently correspond to the lowest bare density of states (indicated by the faintest blue markers), reflecting the combinatorial scarcity of macroscopic configurations with maximal double occupancies. 
 (b) Convergence of the dynamically predicted bias-corrected variance $\sigma_t^2$ toward the effective target variance $\sigma_{\mathrm{target,eff}}^2$.  
(c) Logarithmic decay of the variance mismatch $|\sigma_{\mathrm{target,eff}}^2 - \sigma_t^2|$ during the feedback iteration. 
    \rev{The effective target variance (red dashed line) is computed from the trace invariants of the full Hamiltonian and then adjusted by subtracting the approximate kinetic broadening term (requiring only $\mathcal{O}(D)$ sparse operations rather than $\mathcal{O}(D^3)$ eigendecomposition), ensuring compatibility with the diagonal hidden-variable basis.} All energies are expressed in units of the hopping amplitude $J$.
}
	\end{figure}

	\begin{figure}
		\centering
\includegraphics[width=0.9\linewidth]{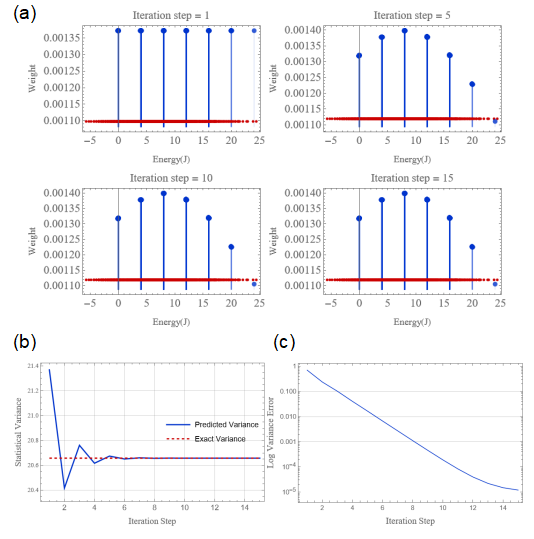}
		\caption{
Same as Fig.~1, but for $L=6$, $N_{\max}=2$, $J=1.0$, $U=4.0$, $F=0.8$, learning rate $\eta=0.8$,
and $15$ iterations.
The larger learning rate produces damped oscillatory convergence before the variance locks onto the target value.
}
	\end{figure}


Fig.1 demonstrates the performance of Algorithm~1 in the overdamped regime. 
The parameters are chosen as $L=4$, $N_{\max}=2$, $J=1.0$, $U=4.0$, $F=0.8$, 
$\eta=0.1$, and the total number of feedback iterations is $80$. The hidden 
interaction energies are defined as
\rev{$M_i=\sum_{\ell=1}^{L}\frac{U}{2}n_{i,\ell}(n_{i,\ell}-1)$ with $n_{i,\ell}$ the particle number in the $i$-th Fock configuration at site $\ell$,}
with the weights $w_i$ dynamically updated by the thermodynamic feedback loop.
Figure~1(a) shows the evolution of the learned weights over the hidden variables 
$M_i$. The red markers denote the exact eigenvalues of the full driven Bose-Hubbard Hamiltonian, while the blue markers denote the predicted weights assigned to the diagonal interaction energies. Initially, the weights are nearly uniform; during the 
iteration process, the feedback mechanism redistributes the weights and generates a 
smooth envelope over the interaction-energy manifold. This smooth envelope visually captures how the localized discrete energy levels are hybridized by the complex interplay among the coherent driving $F$, hopping $J$, and interaction $U$.


The feedback potential modulates the probability distribution across the Hilbert space based on the instantaneous variance mismatch $\Delta_{\sigma^2}$.
For current variance $\Delta_{\sigma^2} > 0$, the algorithm applies an inverted Gaussian potential (positive parabola). This potential is maximal at the spectral edges (far from the mean $\mu_t$), amplifying the weights of the tail states via $w_{new} \propto w_{old} e^{+E_i}$, thereby expanding the distribution width.
When the current variance exceeds the target ($\Delta_{\sigma^2} < 0$), a confining harmonic potential (negative parabola) is applied. This imposes a strong negative feedback $E_i \ll 0$ at the boundaries exponentially suppressing the tail weights via $w_{new} \propto w_{old} e^{-|E_i|}$ and forcing the distribution to contract towards the mean, \rev{reducing the variance until the variance-matching condition is restored.}

Driven by the adaptive feedback mechanism defined in Eqs.\ref{eq:feedback}–\ref{18}, the dynamic transition between expansive heating and confining cooling automatically reshapes the distribution. 
Consequently, the algorithm constructs a coarse-grained spectral envelope (the blue points in Figs.1-2) that captures the target macroscopic width despite the intrinsic differences in the underlying microscopic basis.

\w{The macroscopic degeneracy $\Omega$ of a specific occupation profile $\{m_k\}$ is strictly governed by the multinomial distribution $\Omega(\{m_k\}) = L! / \prod_k m_k!$. If without driving, an increase in empty sites ($m_0$) is necessary to accommodate the localized clusters, thus the rapid growth of the $m_0!$ term in the denominator rigorously dictates an exponential collapse in the bare local density of states $\rho_{bare}(E)= \sum_{i=1}^{D} \delta(E - M_i) = \sum_{m_2} \Omega(m_2) \delta(E - m_2 U)$. 
The effective level density $\rho(E) = \sum_i w_i \delta(E - M_i)$ inherently respects this underlying Hilbert-space structure. Because there are exponentially fewer computational basis vectors available to support these extreme energies, the algorithm intrinsically assigns lower cumulative weights (populated states) to this region without requiring an explicit energetic cutoff. This exponential decay in phase-space volume explains the appearance of the faintest blue markers at the right edge of the spectral scatter plot in Fig. 1(a).
In our case, the energy reads $M_i= \sum_{\ell=1}^{L} \frac{U}{2} n_{i,\ell}(n_{i,\ell} - 1) = m_2 U$,
thus for a fixed $m_2$ the combinatorial degeneracy is $\Omega=\sum_{m_0+m_1=L-m_2} \Omega(\{m_k\}) = \binom{L}{m_2} \sum_{k=0}^{L-m_2} \binom{L-m_2}{k}=\binom{L}{m_2} 2^{L-m_2}$.
For $L=4$, the interaction energy is quantized by the number of doubly occupied sites $m_2$, such that $M_i = m_2 U$. Intermediate-energy manifolds are highly degenerate (e.g., $32$ distinct microstates for $M_i = U$), while the local bare density of states at high energy tail ($M_i = 4U$) is overwhelmingly suppressed, supported by only a single uniquely clustered microstate $|2,2,2,2\rangle$ (smallest number of distinct permutations). 
}

Fig.1(b) shows the convergence of the weighted predicted variance toward the effective target value. The red dashed line denotes the effective target variance $\sigma_{\mathrm{target,eff}}^2$
 obtained from the exact full-spectrum variance after subtracting an approximate kinetic broadening term. 
The blue solid line tracks the dynamically predicted bias-corrected variance $\sigma_t^2$. 
For the present parameter choice, the predicted variance approaches the effective target monotonically from above, which is characteristic of an overdamped relaxation.
This monotonically decreasing behavior confirms the stability of our loss function optimization.

\rev{Through this iterative optimization, Algorithm~1 reconstructs a coarse-grained spectral envelope whose bias-corrected second moment matches the operational effective target $\sigma_{\mathrm{target,eff}}^2$. The distribution dynamically expands or contracts until this macroscopic variance-matching condition is reached, to form a stable envelope that the predicted
weights cover the full energy range of the
exact spectrum and reproduce the target ensemble. Fig.~1(c) quantifies the precision of the convergence: on the logarithmic scale, the mismatch $|\sigma_t^2-\sigma_{\mathrm{target,eff}}^2|$ descends linearly, demonstrating an exponential decay that ultimately
reaches a magnitude below $10^{-3}$. This demonstrates stable macroscopic moment matching. The local GOE-like surrogate spacing statistics are not inferred from this variance convergence alone; they are examined separately using Algorithm~2 and benchmarked against exact diagonalization.}

\w{In our model, the maximum number of particles per site is truncated to $N_{max} = 2$. Because the driving term $F(\hat{a}^\dagger_1 + \hat{a}_1)$ explicitly breaks global particle number conservation, the system can explore all possible combinations of these local states across the lattice.
The total dimension of the Hilbert space in the present truncated basis is $D=(N_{max} + 1)^L$, which is also the number of hidden-variable states and Fock states in the truncated basis.}
In Fig.1 we set the learning rate as $\eta=0.1$ and $L=4$ which is of the overdamped regime.
The total dimension of the Hilbert space is $3^4= 81$.
Since the system reacts slowly to the error, the prediction approaches the target variance slowly and monotonically.
\rev{The blue envelopes and the convergence curves show that the deterministic thermodynamic feedback reconstructs a smooth probability measure whose bias-corrected variance approaches the operational effective target. The broad support of the reconstructed weights indicates extensive participation in the hidden-variable measure, but is not by itself taken as proof of microscopic eigenstate ergodicity. The independent direct-diagonalization result $\langle r\rangle_{\mathrm{ED}}=0.5389$, together with the eigenstate-IPR scaling discussed below, provides the microscopic benchmark for GOE-like chaotic delocalization.}

In overdamped regime,
\rev{the algorithm exhibits a smooth and monotonic convergence of the predicted variance $\sigma_t^2$ toward the effective target $\sigma_{\mathrm{target,eff}}^2$. The smaller learning rate suppresses overshooting and produces an overdamped response.}
\w{The self-attention weights dynamically evolve into a stable, continuous envelope that accurately reproduces the macroscopic spectral variance of the exact Hamiltonian. This broad distribution of probability weights across multiple interaction energy scales demonstrates strong Hilbert space delocalization. Such extensive mixing in the Fock space serves as a fundamental prerequisite for the emergence of quantum chaos.}

Figure~2 shows the same reconstruction protocol for a larger system size and a larger learning rate, with parameters $L=6$, $N_{\max}=2$, $J=1.0$, $U=4.0$, $F=0.8$, $\eta=0.8$, and $15$ iterations. In contrast to Fig.~1, the larger learning rate induces an underdamped response: the weighted variance overshoots the target value, then oscillates with decreasing amplitude before converging. This behavior illustrates that the learning rate mainly controls the stiffness and transient dynamics of the feedback loop, rather than the final variance-matching objective itself.
Higher learning rate results in significant oscillation,
and the distribution explodes outward instantly, making the predicted variance higher than the target variance.
Then the error becomes negative. The algorithm applies massive cooling,
and the variance collapses below the target again.
\rev{The trajectory exhibits a rapidly damped oscillation before settling near the effective target variance, illustrating the transient sensitivity to the learning rate.}

\section{Quantitative Discrimination of Wigner-Dyson Statistics}

\w{
In conventional Wigner-Dyson statistics, the eigenvalues $E_i$ play the role of dynamical degrees of freedom as a function of off-diagonal coupling $J$ and $F$.
\rev{In our self-attention-inspired framework, we do not directly calculate the shifting eigenvalues.} 
Instead, the basis energies (the hidden variables $M_i$) are fixed on a discrete grid, and the algorithm dynamically optimizes the probability weights $w_i$ assigned to these states to meet the overall macroscopic variance. 
Thus all the statistical fluctuations that characterize quantum chaos are transferred from the energy domain into the weight domain.
}

\w{
}

To verify whether the predicted spectral weights $\mathbf{W}$ successfully reconstruct the quantum chaotic signature, we introduce a discrimination algorithm based on the nearest neighbor spacing distribution (NNSD) and the Dyson index $\beta$.

The hallmark of quantum chaos is the repulsion between energy levels, described by the Wigner-Dyson surmise\cite{Mehta}
\begin{equation}
	P(s) = a_\beta s^\beta \exp(-b_\beta s^2),
\end{equation}
where $s_i = \frac{w_i}{\langle w_i \rangle} = \frac{w_i}{1/D} = D \cdot w_i$ is the normalized statistical level spacing. 
In a weighted spectrum, a state with a very high weight $w_i$ occupies a large volume in the probability space, which is equivalent to having a large spacing in the cumulative distribution function.

\w{To compare our reconstructed spectrum with conventional Wigner--Dyson statistics, we perform an unfolding procedure in the cumulative probability space. In our algorithmic framework, the basis energies (hidden variables $M_i$) are fixed, and the dynamical degrees of freedom are entirely encoded in the spectral weights $w_i$. For an ascending ordered set of interaction energies $M_1 < M_2 < \dots < M_D$, we define the effective cumulative density of states as the discrete cumulative sum $x(M_i) = \sum_{j=1}^{i} w_j$. 
Thus the effective spacing between adjacent states in this unfolded cumulative probability space is given directly by the weight itself: $\Delta x_i = x(M_i) - x(M_{i-1}) = w_i$. By normalizing this effective spacing by its ensemble average $\langle w \rangle = 1/D$, we immediately arrive at the normalized surrogate spacing $s_i = w_i / \langle w \rangle = D \cdot w_i$, as defined above. 
In this weighted spectrum, a state with a very high weight $w_i$ occupies a large volume in the probability space, which effectively corresponds to a large surrogate level spacing in the cumulative distribution function. \rev{Therefore, the normalized weights define a surrogate spacing distribution in cumulative probability space. Its comparison with the GOE surmise is used as a phenomenological diagnostic; it is not an identity with the exact eigenvalue-spacing distribution. The independent ED analysis below tests whether the surrogate and eigenvalue-based diagnostics exhibit the same trend toward GOE-like level repulsion.}}

\w{
The basis energies (hidden variables $M_i$) are fixed here, and the dynamical degrees of freedom are the spectral weights $w_i$. 
The effective spacing between adjacent states in this cumulative space is given by the difference $\Delta x_i = x_i - x_{i-1} = w_i$.
In this weighted spectrum, a state with a very high weight $w_i$ occupies a large volume in the probability space, which effectively corresponds to a large surrogate spacing in the cumulative distribution function. 
}

For integrable systems with Poisson distribution, $\beta=0$ and $P(s) \sim e^{-s}$. For chaotic systems, $\beta=1$ (GOE), $\beta=2$ (GUE), or $\beta=4$ (GSE). Higher-order extensions of Wigner--Dyson spacing statistics have also been studied in random-matrix ensembles, providing a broader theoretical context for spectral-correlation diagnostics beyond nearest-neighbor statistics \cite{Rao2020}.
$P(s) \to 0$ as $s \to 0$ is a signal of level repulsion unlike the case of localization.
\rev{The deterministic feedback smooths the coarse-grained weights, whereas the stochastic step restores local fluctuations. In the GOE-like regime, the resulting surrogate spacing distribution suppresses near-zero spacings relative to the Poisson reference.}
\rev{The normalized weights therefore generate the surrogate distribution $P_{\mathrm{sur}}(s)$ used in this reconstruction.} 


\w{In Algorithm 2, we apply the log-normal multiplicative perturbations (additive Gaussian noise in log space), to preserve the positivity of weights and restore local fluctuations around the smooth thermodynamic envelope (produced by Algorithm 1).
The dynamical effect of this stochastic noise injection is illustrated in Fig.3. As shown in the spectral snapshots, the multiplicative perturbation restores the local microscopic roughness of the weights. Because we employ an overdamped learning rate ($\eta=0.1$) for this noisy reconstruction, the predicted variance remains stably locked to the effective target variance without being disrupted by the local fluctuations. This confirms that our dual-architecture framework can independently control the macroscopic envelope and the microscopic spectral roughness.
}
For weighted variance smaller than the target value, the heating process broadens the weight distribution by enhancing contributions farther from the weighted mean, thereby increasing the variance. \rev{When the current variance exceeds the target value, the cooling mechanism concentrates the weights closer to the mean and reduces the variance.}
While this achieves the global variance target, it wipes out the local fluctuations that define Wigner-Dyson statistics.
\w{
Since the underlying driven Bose-Hubbard Hamiltonian is real-symmetric, preserving time-reversal symmetry, the chaotic regime is of the GOE class. The purely deterministic feedback (Algorithm 1, $\sigma_{\text{noise}}=0$) yields an artificially smooth effective measure, which fails to capture the intrinsic local level repulsion. 
Without the application of local stochastic roughening, the unfolded spacing $s_i = w_i/\langle w \rangle$ clusters tightly around 1. This leads to \rev{a highly concentrated surrogate-spacing distribution} $P(s)$, as shown in Fig.4(c).
To recover the microscopic Wigner-Dyson statistics, the local stochastic fluctuation must be injected, which smoothly transitions the spacing distribution to the correct GOE limit as depicted in Fig.4(d).
Here in Algorithm 2 the phenomenological noise amplitude $\sigma_{noise}$ acts as an effective Langevin fluctuation strength,
\rev{For the noisy realization shown in Fig.~4(d), the surrogate ratio is evaluated from Eq.~\eqref{eq:surrogate_ratio}: $ \langle r\rangle_{\mathrm{sur}}=0.540339$. Because it is constructed from the reconstructed weights rather than from exact eigenvalues, it is not expected to coincide numerically with $\langle r\rangle_{\mathrm{ED}}$. The exact $L=6$ ED value, $\langle r\rangle_{\mathrm{ED}}=0.530057$, and the GOE benchmark $\langle r\rangle_{\mathrm{GOE}}\simeq0.5307$ remain the microscopic reference values.}}
This occurs because the dynamically generated weights $w_i$ can become extremely uniform ($w_i \approx 1/D$ with very small variance, yielding $s_i \approx 1$ for all states). In this infinite-temperature limit, the reconstructed spectrum resembles a rigid picket-fence distribution. Such a distribution exhibits anomalously high level repulsion ($\beta \to \infty$) that far exceeds the true GOE chaotic limit, thereby artificially suppressing the microscopic local fluctuations that genuinely characterize quantum chaos.



Thus a stochastic term is necessary to introduce local microscopic noise into the optimization process. This stochastic injection restores the vital local fluctuations, causing the weights $w_i$ to fluctuate organically around the smooth macroscopic mean curve. These localized fluctuations dynamically broaden the unphysical rigid histogram into the correct GOE-like distribution. The effective spacing distribution recovers the Wigner-Dyson surmise for the GOE $    P(s) \approx \frac{\pi}{2} s \exp\left(-\frac{\pi}{4}s^2\right)$,
where the effective spacings linearly repel each other from zero ($P(s) \propto s$ as $s \to 0$) and tend to cluster around the ensemble average ($s=1$).

\rev{Because the reconstruction is formulated directly in cumulative probability space, its normalized surrogate spacings are $s_i=Dw_i$. After ordering the hidden-variable states by $M_i$, we define a purely weight-based adjacent surrogate ratio by
\begin{equation}
 r_{i,\mathrm{sur}}=
 \frac{\min(s_i,s_{i+1})}{\max(s_i,s_{i+1})},
 \qquad
 \langle r\rangle_{\mathrm{sur}}=
 \frac{1}{D-1}\sum_{i=1}^{D-1}r_{i,\mathrm{sur}}.
 \label{eq:surrogate_ratio}
\end{equation}
Equation~\eqref{eq:surrogate_ratio} is eigenvalue-free: it is constructed entirely from the reconstructed weights and does not require the exact gaps $E_{n+1}-E_n$. The arithmetic mean is chosen to mirror the standard ED definition of the mean adjacent-gap ratio. In particular, we do not apply an additional weight factor to the average, because the weights have already entered the statistic through $s_i=Dw_i$; weighting the average a second time would preferentially emphasize large-$w_i$ regions and would define a different, more biased diagnostic. We use $P_{\mathrm{sur}}(s)$ and $\langle r\rangle_{\mathrm{sur}}$ only as phenomenological diagnostics of the reconstructed GOE-like trend. They are kept distinct from the exact-eigenvalue ratio $\langle r\rangle_{\mathrm{ED}}$ defined below, which provides the independent microscopic benchmark. The distribution shape can additionally be compared with the standard Wigner--Dyson $\beta$-ensembles through the Kullback--Leibler (KL) divergence.}

A global unfolding is realized by defining the normalized spacing as $s_i = w_i / \langle w \rangle$ because the Algorithm 2 forces the weights to form a smooth envelope that matches the target statistics, the "local mean weight" is effectively handled by the thermodynamic feedback. Dividing by the global mean $\langle w \rangle$ is sufficient to unfold the spectrum for statistical analysis in this specific context.
Since the spectral weights $w_i$ directly represent the local phase space volume, the unfolding procedure simplifies to normalizing the weights by their ensemble average: $s_i = w_i / \langle w \rangle$. This maps the weight distribution directly onto the dimensionless level spacing statistics.

\w{To further validate the ratio statistic used in our weighted-spectrum approach,
we benchmark it against direct diagonalization (ED) of the Bose--Hubbard Hamiltonian
for several accessible system sizes. In the ED calculation, the eigenvalues are first
sorted in ascending order, and the adjacent gaps are defined as
\begin{equation}
\delta_n = E_{n+1}-E_n.
\end{equation}
We then compute the standard gap-ratio statistic
\begin{equation}
r_n=\frac{\min(\delta_n,\delta_{n+1})}{\max(\delta_n,\delta_{n+1})},
\qquad
\langle r\rangle_{\mathrm{ED}}=\frac{1}{N_r}\sum_n r_n.
\label{eq:standard_ratio_ED}
\end{equation}
\rev{The resulting values of $\langle r\rangle_{\mathrm{ED}}$ are compared with the universal benchmarks}
for Poisson, GOE, and GUE statistics. This comparison provides an independent check
that the weighted-spectrum prediction reproduces the same trend toward level repulsion
as the conventional eigenvalue-based analysis.
For the present Hamiltonian, which is real symmetric for real $J,U,F$, the
relevant chaotic benchmark is GOE rather than GUE. For example, at $L=4$ we
obtain from exact diagonalization
$\langle r\rangle_{\mathrm{ED}} = 0.5389$,
which is very close to the GOE prediction and clearly distinct from the
Poisson value. This provides an independent check that the weighted-spectrum
prediction reproduces the same trend toward level repulsion as the conventional
eigenvalue-based analysis.}

\w{To validate the chaotic nature of the target Hamiltonian and provide an independent benchmark for our weighted-spectrum prediction, we performed exact diagonalization (ED) across various system sizes ($L = 4$ to $8$). 
}

\w{
Table I presents the mean adjacent-gap ratio $\langle r \rangle_{ED}$ obtained via direct exact diagonalization for increasing system sizes from $L=4$ to $L=8$. The results show that while finite-size fluctuations exist for small size $L=5$, the ratio rapidly stabilizes towards the GOE benchmark value $\langle r \rangle_{GOE} \approx 0.5307$ for larger systems ($L\ge 6$). 
The deviation at $L=5$ where $\langle r\rangle_{\mathrm{ED}}\approx 0.499$ is attributed to finite-size effects, which are significant even in a small Hilbert space.
\rev{These exact-diagonalization results provide an independent eigenvalue-based benchmark for the underlying Hamiltonian and are independent of algorithmic hyperparameters such as the learning rate used in the attention-inspired reconstruction.} 
\rev{The GOE-like suppression of small surrogate spacings is qualitatively consistent with the level-repulsion trend established independently by the exact eigenvalue analysis; we do not identify the surrogate statistic with the exact ED ratio.}
}

\begin{table*}[t]
\centering
\caption{Mean adjacent-gap ratio $\langle r\rangle$ and IPR from exact diagonalization and algorithmic weights for several system sizes $L$.  The reference values $1/D$ and $3/D$ are shown for comparison. \rev{The column $\mathrm{IPR}_w$ is evaluated from the converged deterministic Algorithm~1 weights before the stochastic roughening of Algorithm~2; therefore the entries in this table do not depend on $\sigma_{\mathrm{noise}}$.}}
\begin{tabular}{c c c c c c c}
\hline
$L$ & $D$ & $\langle r\rangle_{\mathrm{ED}}$ & IPR$_{w}$  & $1/D$ & $\langle {\rm IPR}_{e}\rangle$ & 3/D\\
\hline
4 & 81  & 0.538893 & 0.0123617  & 0.0123457 & 0.0991362 &0.037037\\
5 & 243 & 0.499171 & 0.00411876  & 0.00411523 & 0.0430899 & 0.0123457\\
6 & 729 & 0.530057 & 0.0013726 & 0.00137174 & 0.0159904& 0.00411523\\
7 & 2187& 0.527335 & 0.000457454 & 0.000457247 & 0.00548612 & 0.00137174\\
8 & 6561& 0.530133 & 0.000152469  & 0.000152416 & 0.00180517 & 0.000457247\\
\hline
\end{tabular}
\label{tab:benchmark}
\end{table*}

\w{
}

\w{\rev{In the bosonic occupation-number basis $|n\rangle=|n_{1},n_{2},\cdots,n_{L}\rangle$ with truncation $0\le n_{\ell}\le N_{\max}$, the boundary drive removes the restriction to a fixed total-particle-number sector. We treat these Fock states as the basic input configurations for the attention-inspired reconstruction and use the exact eigenstate IPR as an independent diagnostic of Hilbert-space delocalization.}
As shown in Table~\ref{tab:benchmark},
the mean eigenstate IPR $\langle{\rm IPR}_{e}\rangle=\frac{1}{D}\sum_{\alpha=1}^{D}\sum_{n=1}^{D}|\psi_{\alpha}(n)|^4$ in the Fock basis decreases as $\mathcal O(D^{-1})$ with increasing system size, 
indicating that the eigenstates are extensively delocalized over a finite fraction of the Hilbert space in Fock basis,
although they remain above the ideal GOE value $3/D$ due to finite-size and basis-structure effects.
\rev{Figure~\ref{fig:ipr_scaling} displays the finite-size scaling on a $\log_{10}D$--$\log_{10}\langle\mathrm{IPR}_e\rangle$ scale, which resolves the larger-system points more clearly than a linear $1/D$ axis. A power-law fit to the stabilized larger-system points $L=6,7,8$ gives $\langle\mathrm{IPR}_e\rangle\propto D^{-\alpha}$ with $\alpha=0.992768$ and $R^2=0.999878$, consistent with asymptotic $D^{-1}$ scaling. The equivalent direct linear fit in $1/D$ is $\langle\mathrm{IPR}_e\rangle=11.5997/D+9.93375\times10^{-5}$ with $R^2=0.999897$. Thus both representations support the same $D^{-1}$ finite-size trend; the log--log representation is used in Fig.~\ref{fig:ipr_scaling} only to make the scaling exponent and the separation of the data points visually explicit. We do not infer the finite-size prefactor to be exactly the ideal GOE value $3/D$.}
While the weight IPR $\mathrm{IPR}_w=\sum_{i=1}^{D} w_i^2$ becomes increasingly close to its lower bound $1/D$ with increasing system size (the set $\{w_i\}$ is not concentrated on only a few states but spread across many states), 
thus the algorithmic probability mass remains broadly distributed over an extensive number of hidden-variable states as the system grows,
which indicates that Algorithm~1 reconstructs an almost maximally spread (uniform) smooth spectral envelope over the $D$ hidden-variable states. 
\rev{Thus, in the GOE-like chaotic regime, the attention-inspired reconstruction captures the coarse-grained spreading of the spectral measure, while the local surrogate-spacing fluctuations are supplied phenomenologically by Algorithm~2.}
For finite system sizes, the measured $\langle\mathrm{IPR}_e\rangle$ is  larger than the ideal GOE reference value $3/D$, since local connectivity constraints, basis-dependent structure, and finite-size effects prevent a perfectly featureless weight distribution.
Together with the adjacent-gap ratio approaching the GOE benchmark,
the systematic decrease of $\langle\mathrm{IPR}_e\rangle$ with increasing system size provides additional evidence of the GOE-like chaotic delocalization in Fock basis. 
\rev{Thus the weight-based thermodynamic feedback scheme reconstructs the coarse-grained spectral envelope through global moment matching without needing to resolve individual eigenstate components.}}


The scaling behavior $\langle IPR_e \rangle \sim \mathcal{O}(D^{-0.993})$ provides microscopic evidence that the typical eigenstates of the strongly interacting driven Bose-Hubbard chain are extensively delocalized across the many-body Fock space. In the strongly interacting and ergodic regime, typical many-body eigenstates are no longer localized but spread out widely, being composed of a macroscopic number of unperturbed Fock states close in energy\cite{Neuenhahn,Flambaum}.
\rev{The coherent driving term $F(\hat{a}_1^\dagger + \hat{a}_1)$ explicitly breaks global particle-number conservation and connects otherwise separated particle-number sectors. Together with hopping, it enables hybridization across a much larger portion of the truncated Fock space. The observed decrease of $\langle\mathrm{IPR}_e\rangle$ and the GOE-like adjacent-gap ratios are therefore consistent with extensive Fock-space delocalization in the parameter regime studied here.} 

Since extensively delocalized eigenstates are superpositions of a macroscopic number of bare Fock states $\mathcal{O}(D)$. This massive superposition allows the law of large numbers to suppress the eigenstate-to-eigenstate fluctuations of few-body observables, and the variance of fluctuations (Fock-space weights) decay proportionally to the mean IPR
or statistical variance of the state's probability distribution over the bosonic Fock basis\cite{Neuenhahn}. \rev{Thus the observed IPR scaling, $\langle \mathrm{IPR}_e\rangle\sim e^{-(0.993\ln3)L}$ for the fitted larger sizes, is consistent with increasingly extensive Fock-space delocalization.}



While the emergent Wigner-Dyson statistics suggest a transition to quantum chaos, the underlying many-body eigenstates carry robust correlations inherent to the two-body interaction such that the components of compound states in many-body systems are not purely independent Gaussian variables.
\rev{Random two-body-interaction models provide useful qualitative context for the fact that residual correlations can persist among components of broadly delocalized many-body eigenstates, so a few-body Hamiltonian need not behave as a structureless full random matrix. We use this observation only as motivation for retaining a structured Fock-space description. The stochastic roughening in Algorithm~2 is a phenomenological operation on the surrogate spectral measure.
}
Because of the residual correlations, different Fock components (eigenstate amplitude) are not independent Gaussian variables (as demonstrated in random two-body interaction models). 
In the few-body structured Hamiltonian,
each basis state is only connected to a restricted set of nearby configurations through few-body processes, unlike the structureless full random matrix.
\rev{The fitted exponent $\alpha\simeq0.993$ demonstrates extensive delocalization despite the few-body structure of the Hamiltonian. It does not by itself determine the magnitude of residual correlations between individual eigenvector components. As the accessible truncated Fock space $D=3^L$ (number of active local degrees of freedom) grows, such residual correlations may persist because the Hamiltonian remains an embedded few-body operator rather than a structureless full GOE matrix; the number of independent one- and two-body parameters is much smaller than the total number of many-body matrix elements so the eigenvector components inherit structured correlations\cite{Flambaum}.} 

\w{
It is crucial to note the basis dependence when discussing interaction-induced ergodicity. In momentum-space fermionic models\cite{Neuenhahn}, the interaction acts as off-diagonal perturbation in the noninteracting Fock basis, causing typical eigenstates to spread out widely over nearby Fock configurations. While in our real-space model, the on-site interaction $U$ is diagonal in the occupation-number Fock basis. Thus in limit $U/J \to \infty$, there is Fock-space localization (analogous to the Mott-insulator limit) characterized by a high IPR.
The Fock-space delocalization emerges in the parameter regime in which the combined action of hopping $J$, drive $F$, and interaction $U$ efficiently mixes a large number of nearby bosonic configurations, where the eigenstates become (structured) delocalized in the Fock basis,
whereas the extreme limits ($U=0$ or $U \to \infty$) would trivially reduce to localized or integrable dynamics.
Thus the two-body nature of $U$ ensures that these broadly delocalized components are not merely independent Gaussian variables within a smooth energy-shell envelope but carry residual correlations even in strongly mixed regimes, and thus enhance the fluctuation.
Such residual correlations, inherited from the one- and two-body structure, can survive and contribute to observable fluctuations, thus existing even in chaotic regime. In realistic chaotic many-body systems, the components can be nearly Gaussian at the coarse-grained level and still carry nontrivial correlations because the Hamiltonian is built from few-body operators rather than a structureless full GOE matrix. 
}

\begin{algorithm}[H]
	\caption{Statistical Discrimination via $P(s)$ with Noise Injection}
	\label{alg:discrimination}
	\DontPrintSemicolon
	\KwIn{Predicted Weights $\mathbf{W}$, Interaction Energies $\mathbf{M}$, \textbf{Flag} `EnableNoise`, Noise strength $\sigma_{noise}$.}
	\KwOut{Effective Dyson Index $\beta_{en}$, Phase Classification.}
	

\textbf{Step 1: Stochastic Fluctuation Injection}\\
\If{EnableNoise is True then}
{\Comment{Restore local multiplicative fluctuations around the smooth envelope}
Generate Gaussian log-noise: $\xi\sim \mathcal{N}(0,\sigma_{\mathrm{noise}}^2)$\;
Update weights: $w_i \leftarrow w_i \exp(\xi_i)$\;
Renormalize: $W \leftarrow W / \sum_j W_j$\;}

	\BlankLine
	\textbf{Step 2: Spectrum Unfolding for $P(s)$} \\
	\rev{Sort energies and weights: $(M_1,w_1),\dots,(M_D,w_D)$}\;
	\rev{Compute the weighted cumulative function $x(E)=\sum_{i=1}^{D}w_i\Theta(E-M_i)$}\;
	\rev{Transform to cumulative surrogate spacings: $s_i=x(M_i)-x(M_{i-1})=w_i$}\;
	Normalize spacings (Mean level spacing $\langle s \rangle = 1$): 
	$\tilde{s}_i = s_i / \langle s \rangle$\;
	
	\BlankLine
	\textbf{Step 3: $\beta$-Ensemble Fitting via $P(s)$} \\
	Define theoretical Wigner-Dyson distributions $P_{WD}(s, \beta)$:\\
	$\beta=0$ (Poisson): $P(s)=e^{-s}$\\
	\rev{$\beta=1$ (GOE): $P(s)=\frac{\pi}{2}s\exp[-\pi s^2/4]$}\\
	$\beta=2$ (GUE): $P(s) = \frac{32}{\pi^2}s^2 e^{-\frac{4}{\pi}s^2}$\\
	Construct histogram $H(\tilde{s})$ of normalized spacings $\tilde{s}_i$\;
	
	\Comment{Calculate Kullback-Leibler Divergence}
	\For{$\beta \in \{0, 1, 2, 4\}$}{
		$D_{KL}(\beta) = \sum H(s) \ln \frac{H(s)}{P_{WD}(s, \beta)}$\;
	}
	Identify $\beta_{en} = \arg\min_\beta D_{KL}(\beta)$\;
	
	\BlankLine
	\textbf{Step 4: Classification} \\
	\uIf{$\beta_{en} == 0$}{
		Phase $\leftarrow$ \textbf{Localized (Poisson)}
	}
	\uElseIf{$\beta_{en} == 1$}{
		Phase $\leftarrow$ \textbf{Chaotic (GOE)}
	}
	\uElseIf{$\beta_{en} == 2$}{
		Phase $\leftarrow$ \textbf{Chaotic (GUE)}
	}
	\Else{
		Phase $\leftarrow$ \textbf{Chaotic (GSE)}
	}
	
	\Return $\beta_{en}, \text{Phase}$\;
\end{algorithm}


\section{Conclusion}

\rev{We have developed an attention-inspired spectral prediction algorithm that reconstructs coarse-grained statistical properties of the driven Bose-Hubbard model without requiring a full eigendecomposition inside the reconstruction loop. This reduces the numerical overhead relative to full-spectrum diagonalization but does not remove the exponential growth of the truncated Hilbert-space dimension with $L$.} By mapping the interaction-induced energy fluctuations to a set of modulable hidden variables, we introduced a deterministic thermodynamic feedback mechanism, comprising inverted Gaussian and confining potentials, that dynamically drives the macroscopic spectral weight envelope toward the exact statistical moments.



\rev{The coherent boundary driving field $F$ plays a non-perturbative role. It breaks the global $U(1)$ particle-number conservation and connects otherwise separated particle-number sectors. Together with the hopping $J$, the drive enhances configuration hybridization against the localization tendency of the interaction-dominated limit. This broad participation is reflected by the high Shannon entropy of the reconstructed weight distribution and by the nontrivial particle-number covariance structure, while the exact ED benchmarks independently establish GOE-like spectral statistics for the accessible system sizes.}

\rev{The emergence of GOE-like surrogate spacing statistics in the strongly mixed regime demonstrates that, while the deterministic thermodynamic feedback accurately captures the macroscopic spectral envelope, the subsequent injection of multiplicative log-normal stochastic noise is crucial for restoring local fluctuations in the reconstructed measure.} 
\rev{At the eigenstate level, further validation could combine the present spectral diagnostics with ETH-style tests of diagonal and off-diagonal matrix elements, as used recently in bosonic many-body chaos studies\cite{Prashant2026}. Such tests require microscopic eigenvectors and are therefore complementary to, rather than part of, the present coarse-grained reconstruction.}


\appendix

\section{Renormalization Group Analysis of Spectral Variance}
\label{app:rg}
\label{app:RG_variance}

In this appendix, we provide the theoretical justification for the variance stability observed in our algorithm. We treat the weight optimization process as a discrete dynamical system flowing towards a fixed point under a renormalization group (RG) transformation.

Consider a binary subgroup of Fock states with weights $\{w_1, w_2\}$ satisfying the local normalization constraint $w_1 + w_2 = 1$. The statistical fluctuations of this subgroup are characterized by the sum of squared weights, which relates to the product term (local IPR)
\begin{equation}
	\begin{aligned}
		\sum_{i=1}^2 w_{i}^2 = w_1^2 + (1-w_1)^2 = 1 - 2w_{1}(1-w_{1}).
	\end{aligned}
\end{equation}
We define the dynamical scaling function $f_{\chi} \equiv 2w_{1}(1-w_{1})$. The maximum entropy state (uniform distribution $w_1=w_2=1/2$) corresponds to $f_\chi = 1/2$.
 By associating the iterative optimization steps with a continuous flow parameter $\chi$ (representing the effective Hilbert space scale), the fixed point of the chaotic flow is strictly defined by the saturation condition
\begin{equation}
	\begin{aligned}
		\label{eq:fixed_point}
\lim_{\chi\rightarrow\infty}f_{\chi}=\frac{1}{2}.
	\end{aligned}
\end{equation}
In the vicinity of the fixed point, the flow equation governing the dynamically predicted variance $\sigma^2$ can be derived. 
\w{The strong hybridization induced by the driving field $F$ and the interaction $U$ renormalizes the effective sample size of the configurations. Under the RG flow, the dynamical response of the weight distribution introduces a scale-dependent renormalization factor $Z_\chi \equiv 1 - \partial_\chi f_\chi$.}
By enforcing the invariance of the spectral variance under weight redistribution (which represents the macroscopic conservation of spectral information), we obtain a relation between the function $f_\chi$ and its response to the scale change (the effective thermodynamic variance)
\begin{equation}
	\begin{aligned}
		\sigma^2 \sim \frac{f_{\chi}}{Z_\chi} =
        \frac{f_{\chi}}{1-\partial_{\chi}f_{\chi}}.
	\end{aligned}
\end{equation}

The condition for the algorithm to settle into a Wigner-Dyson distribution is equivalent to the vanishing of the beta-function for the macroscopic variance
\begin{equation}
	\begin{aligned}
		\beta_{\sigma} = \partial_\chi \left( \frac{f_\chi}{1-\partial_\chi f_\chi} \right) \to 0 \quad \text{as} \quad \chi \to \infty.
	\end{aligned}
\end{equation}
Because the flow stalls at the fixed point ($\partial_\chi f_\chi \to 0$), the renormalization factor approaches unity ($Z_\chi \to 1$), ensuring that the effective variance stabilizes exactly at $\sigma^2 \sim 1/2$. 
This guarantees that the dynamically predicted weights $w_i$ form a robust macroscopic envelope that capture the level fluctuations of the chaotic many-body spectrum.


\section{Mapping to $\beta$-Ensembles via Tridiagonal Matrices}
\label{app:beta_ensemble}

As a complementary diagnostic beyond the variance-matching criterion, we also characterize the reconstructed spectrum using the Dyson index \(\beta_{en}\) ($\beta_{en}=1,2,4$ for GOE, GUE, and GSE ensembles).
The joint probability density function (JPDF) for the eigenvalues $\{k_i\}$ in the chaotic regime is given by\cite{Tracy C A,Dumitriu I,Mehta}
\begin{equation} 
	\begin{aligned}
		P(\{k\}) = C_{\beta,N} \prod_{i<j}|k_{i}-k_{j}|^{\beta_{en}} \exp\left(-\frac{1}{2\sigma^2}\sum_{i=1}^{N}(k_{i}-\mu)^2\right),
	\end{aligned}
\end{equation}
where $N$ is the number of ordered levels,
$\sigma^2$ relates to the effective bandwidth of the spectrum, and the normalization parameter reads
\begin{equation}
	\begin{aligned}
		C_{\beta,N}^{-1} = (2\pi \sigma^2)^{N/2} \sigma^{N(N-1)\beta_{en}/2}
\prod_{j=1}^{N}\frac{\Gamma(1+j\frac{\beta_{en}}{2})}{\Gamma(1+\frac{\beta_{en}}{2})}.
	\end{aligned}
\end{equation}

In our algorithm, the initial weight distribution is chosen as a smooth non-singular reference distribution, and the effective Dyson index \(\beta_{en}\) is then inferred dynamically from the reconstructed spacing statistics.
\w{By iteratively adjusting the spectral weights, the algorithm drives the effective repulsion parameter $\beta_{en}$ from the localized Poisson limit ($\beta_{en}\to 0$) toward the GOE chaotic limit ($\beta_{en}\to 1$). In this sense, $\beta_{en}$ is used as a phenomenological diagnostic of the emergence of Wigner-Dyson level repulsion in the reconstructed weighted spectrum.}

\w{By applying the stochastic fluctuation injection to adjust the weights, we lift the fractional ensemble index $\beta_{en}$ in the spectral measure. The coherent driving field $F$ explicitly breaks the global $U(1)$ particle-number conservation, opening resonance channels that hybridize formerly disconnected symmetry sectors. Different to the case that breaking time-reversal symmetry (which would yield a GUE phase), this strong Fock-space mixing drives the system from an integrable localized Poisson distribution ($\beta_{en} \to 0$) toward the fully ergodic GOE limit ($\beta_{en} \to 1$). }

To clarify the relation between the spectral correlations and the eigenvector structure, we employ the Dumitriu-Edelman tridiagonal matrix model \cite{Dumitriu I,Tracy C A}.
For a tridiagonal matrix $T_N$ (representing the Lanczos basis of $H_b$), the Vandermonde determinant (which quantifies level repulsion) of eigenvalues is 
\begin{equation}
	\begin{aligned}
		\prod_{i<j}|k_{i}-k_{j}|^{\beta_{en}}
		= \frac{\prod_{i=1}^{D-1} t_{i}^{\beta_{en} \cdot i}}{\prod_{j=1}^{D} |Q_{j}|^{\beta_{en}}},
        \label{dumitriu}
	\end{aligned}
\end{equation}
where $t_i$ are the off-diagonal elements of the tridiagonal matrix, and $Q_j$ is the first component of the $j$-th eigenvector.
\rev{Equation~(\ref{dumitriu}) provides a formal relation, within the tridiagonal $\beta$-ensemble construction, between the Vandermonde factor and the first components of the corresponding tridiagonal-model eigenvectors. We use this identity only as a complementary random-matrix consistency relation and not as a direct localization measure for the physical Bose--Hubbard eigenstates.} 
\rev{Note that the predicted coarse-grained spectral weights $w_i$ on the hidden variables are distinct from the microscopic eigenvector intensities $|Q_i|^2$. The former define our surrogate spectral measure, whereas the latter characterize individual eigenvectors; no direct identification $w_i=|Q_i|^2$ is made.}
\rev{Accordingly, the tridiagonal construction is used here to motivate the compatibility between spectral repulsion and delocalized eigenvector structure at the ensemble level. In the physical Bose--Hubbard problem, the actual Poisson-to-GOE crossover is established independently by the exact adjacent-gap ratios and the Fock-basis IPR benchmarks, while the reconstructed weights provide the coarse-grained surrogate measure.} 

\rev{Because no finite-$\beta$ flow equation is derived from Eq.~(\ref{dumitriu}), we do not use the tridiagonal identity to infer a quantitative renormalization of $\beta_{en}$ in the physical Bose--Hubbard model. Instead, the Poisson-to-GOE crossover is established from the exact ED benchmarks and from the reconstructed surrogate spacing distribution. In particular, the nearly uniform deterministic weights produce an overly rigid surrogate spacing sequence concentrated near $s=1$, whereas Algorithm~2 phenomenologically restores local fluctuations and yields a GOE-like $P_{\mathrm{sur}}(s)$. This statement concerns the surrogate spectral measure and does not identify the reconstructed weights with the microscopic tridiagonal eigenvector components.}

Thus the Wigner-Dyson statistics in our reconstructed weighted spectrum is consistent with the delocalized chaotic regime identified independently by the exact-diagonalization benchmarks and the decreasing eigenstate IPR. The thermodynamic feedback and stochastic roughening do not directly compute the microscopic eigenvectors of the full Hamiltonian, but provide a phenomenological reconstruction of the same macroscopic trend: as the weights become broadly distributed over the hidden-variable manifold, the effective spacing statistics evolve away from the Poisson limit and toward GOE-like level repulsion. 

\section{\rev{Delocalization and Occupation-Correlation Metrics}}
\label{app:entanglement}

\rev{The covariance and Shannon-entropy diagnostics used below are constructed from the reconstructed diagonal probability weights $\mathbf{W}$ in the occupation-number Fock basis. They quantify occupation-number correlations and participation of the coarse-grained measure. They are not used as direct entanglement witnesses, because the reconstruction does not retain the microscopic many-body phases required for a quantum-state entanglement measure.}

\subsection{Particle Number Covariance Matrix}
 
In our discrete lattice system, we consider the particle number covariance matrix, $C_{ij}$, which quantifies the density-density correlations between different lattice sites $i$ and $j$, induced by the competition between the driving field $F$ and the on-site interaction $U$.

Using the weights $w_k$ assigned to each Fock basis state $|k\rangle = |n_1^{(k)}, n_2^{(k)}, \dots, n_L^{(k)}\rangle$ by the Algorithm 1, the elements of the covariance matrix are 
\begin{equation}
    \begin{aligned}
        C_{ij} &= \langle \hat{n}_i \hat{n}_j \rangle - \langle \hat{n}_i \rangle \langle \hat{n}_j \rangle \\
        &= \left( \sum_{k=1}^{D} w_k  n_i^{(k)} n_j^{(k)} \right) - \left( \sum_{k=1}^{D} w_k n_i^{(k)} \right) \left( \sum_{k=1}^{D} w_k n_j^{(k)} \right),
        \label{C1}
    \end{aligned}
\end{equation}
where $D$ is the dimension of the Hilbert space.
\rev{The diagonal elements ($C_{ii}$) measure local particle-number fluctuations (compressibility). The off-diagonal elements ($C_{ij}$, $i\neq j$) quantify intersite occupation-number correlations in the reconstructed probability measure. Nonzero $C_{ij}$ therefore indicates correlated particle-number fluctuations between different lattice sites. Because the present covariance matrix is reconstructed from diagonal occupation probabilities, $C_{ij}$ alone is not used here as a direct witness of quantum coherence or bipartite entanglement.}

\subsection{\rev{Shannon Entropy as a Measure of Weight Delocalization}}

\rev{To quantify how broadly the reconstructed probability measure is distributed over the truncated Fock basis, we employ the Shannon entropy of the spectral weight distribution. This is a participation/delocalization diagnostic for the coarse-grained weights and is not identified with the quantum entanglement entropy of a microscopic wave function.}
This metric measures the extent of delocalization of the \rev{reconstructed measure} in the interaction basis
\begin{equation}
    \begin{aligned}
        S = - \sum_{k=1}^{D} w_k \ln(w_k).
    \end{aligned}
\end{equation}
\rev{Low entropy ($S\to0$) indicates that the reconstructed weight is concentrated on only a small number of Fock configurations. High entropy ($S\sim\ln D$) indicates that the probability measure is broadly and nearly uniformly distributed across the accessible Fock basis. In the present work this is used as a coarse-grained delocalization diagnostic; it does not by itself establish microscopic entanglement or thermal equilibrium.}

\rev{The particle-number covariance matrix and the Shannon entropy $S=-\sum_i w_i\ln w_i$ provide complementary diagnostics of the reconstructed probability measure: the former characterizes occupation-number correlations, while the latter quantifies its global participation across Fock configurations.}

The calculation of these metrics is integrated into the post-processing phase of our algorithm. The `WeightedStats` module computes the first and second moments of the particle number operators with respect to the learned weights $\mathbf{W}$, allowing for the efficient reconstruction of the full $L \times L$ covariance structure without explicit wavefunction diagonalization.

\w{From our extracted distributions,}
the calculated Shannon entropy is $S \approx 6.59$. Since the simulation was performed on a lattice with $L=6$ sites and local dimension $d_{loc}=N_{max}+1=3$ ($n \in \{0,1,2\}$), the Hilbert space dimension is $D \approx 3^6 = 729$, and the maximum possible entropy is $S_{max} = \ln(729) \approx 6.59$.
\rev{The extracted value $S\approx S_{\max}$ shows that the reconstructed weights are close to the maximally spread probability measure over the accessible basis. Note that this statement concerns the algorithmic participation of the weights instead of 
a direct determination of thermodynamic temperature or microscopic entanglement. For a perfectly uniform measure, $w_i=1/D$ and the Shannon entropy reaches $S_{\max}=-\sum_i \frac{1}{D}\ln(1/D)=\ln D$.}
\rev{Furthermore, the particle-number correlations associated with the reconstructed weights are plotted in Fig.~4.}
\rev{The relatively large diagonal entries indicate substantial local particle-number fluctuations in the reconstructed measure. The off-diagonal entries quantify intersite correlations and may take either sign; negative values indicate occupation-number anti-correlation. The repulsive interaction $U$ contributes to suppressing multiple occupancy, but the detailed covariance pattern reflects the combined effects of $U$, $J$, $F$, and the reconstructed weights. These nonzero off-diagonal covariances are therefore interpreted as occupation-number correlations rather than as a direct proof of quantum coherence.}

When the finite stochastic noise $\sigma_{\mathrm{noise}}$ is turned on (Algorithm 2), it introduces local multiplicative log-normal fluctuations 
to the converged smooth envelope. \rev{For the representative noisy realization used in Fig.~\ref{noise4}(b,d), $\sigma_{\mathrm{noise}}=0.10$ and the reconstructed Shannon entropy is $S=6.48046$. This value remains close to the high-entropy delocalized regime while allowing appreciable local fluctuations in the covariance matrix and surrogate spacing distribution. Because Algorithm~2 is stochastic, realization-dependent quantities fluctuate slightly between runs; all numerical values quoted for Fig.~\ref{noise4} refer to the specific realization displayed there.}

\rev{A perfectly uniform probability distribution ($w_i=1/D$) yields a delta-like surrogate spacing distribution concentrated at $s=1$, rather than a Wigner--Dyson distribution.} The signature of level repulsion ($P(s) \to 0$ as $s \to 0$) only dynamically emerges when this microscopic roughness is injected into the weight landscape. By specifically using log-normal perturbations, we preserve the positivity of the effective spacings while slightly lowering the entropy to restore the essential microscopic fluctuations, thereby avoiding the unphysical divergence at $s \to 0$ that would occur with pure Porter-Thomas ($\chi^2$) noise.

When the stochastic perturbation is introduced in Algorithm 2, the local particle-number variance $C_{ii}$ remains on the same macroscopic order ($\sim 0.66$) because the log-normal fluctuations are generated by an unbiased zero-mean Gaussian noise $\xi \sim \mathcal{N}(0, \sigma_{\mathrm{noise}}^2)$ that is uncorrelated with specific Fock-space occupation patterns. The stochastic roughening does not bias the probability distribution toward clustered or depleted configurations. The reconstructed spectral measure thus maintains its broad participation across the accessible Hilbert space with the microscopic fluctuations to restore GOE-like level repulsion, and minor site-to-site variations in $C_{ii}$ reflecting the boundary-driven non-equilibrium nature.
Because the stochastic perturbation is applied uniformly across the Fock basis, random noise realization introduces subtle site-to-site fluctuations in the reconstructed covariance matrix without destabilizing the overall spectral envelope.

This injection of microscopic fluctuation among the many-body configurations preserves the macroscopic thermodynamic properties of the chaotic phase.
\rev{The stochastic perturbation changes the local structure of the reconstructed measure while leaving the diagonal occupation-number variances of the same overall order. In Fig.~\ref{noise4}, the nearly noise-free reconstruction has $C_{ii}\sim0.66$ and small off-diagonal correlations, while the noisy realization develops stronger site-to-site variations without destroying the broad participation of the weights. These covariance entries are interpreted here as occupation-number fluctuation diagnostics of the reconstructed measure, rather than as a direct transport coefficient or an entanglement witness.}

\rev{A finite stochastic noise $\sigma_{\mathrm{noise}}$ during the optimization slightly reduces the Shannon entropy from its maximum $\ln D$ and supplies the local weight fluctuations required for the GOE-like surrogate spacing profile. The large diagonal values $C_{ii}$ simply show that substantial local particle-number fluctuations remain present after this roughening step, as a hallmark of macroscopic compressibility ($\kappa \propto \sum_{i,j} C_{ij}$) in the chaotic phase.}


	\begin{figure}
		\centering
\includegraphics[width=0.9\linewidth]{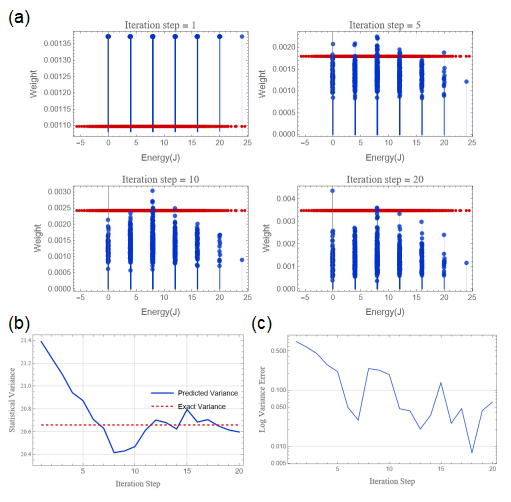}
		\caption{
Simulation results of the noisy spectral reconstruction (Algorithm~1 combined with Algorithm~2) for the driven Bose--Hubbard chain.
\rev{Parameters are $L=6$, $N_{\max}=2$, $J=1.0$, $U=4.0$, $F=0.8$, learning rate $\eta=0.1$, total iterations $=20$, and noise amplitude $\sigma_{\mathrm{noise}}=0.10$.}
The stochastic multiplicative perturbation restores local microscopic roughness around the smooth thermodynamic envelope generated by Algorithm~1.
The predicted macroscopic variance remains stably locked to the effective target variance, while the noisy spectral weights acquire the local fluctuations required for the subsequent reconstruction of Wigner-Dyson statistics.
Compared with the larger-learning-rate case in Fig.2, the smaller learning rate used here leads to a smoother convergence of the macroscopic variance in the presence of noise.}		
	\end{figure}

    	\begin{figure}
		\centering
\includegraphics[width=0.9\linewidth]{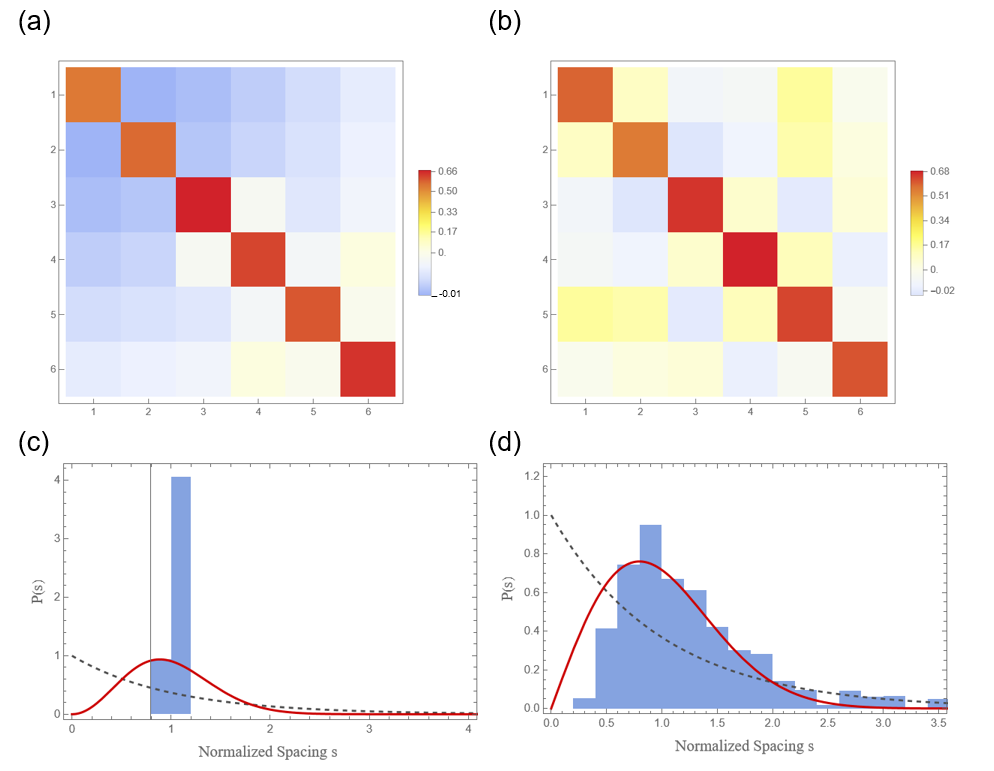}
\caption{
\rev{Emergence of GOE-like surrogate spacing statistics and particle-number correlations in the chaotic regime. The parameters are $L=6$, $N_{\max}=2$, $J=1.0$, $U=4.0$, and $F=0.8$. (a) Particle-number covariance matrix for a nearly noise-free reconstruction with $\sigma_{\mathrm{noise}}=0.001$, for which the numerical covariance range is $[-0.00942066,0.666643]$; the lower color-bar endpoint is displayed as $-0.01$. (b) Covariance matrix for the representative noisy reconstruction with $\sigma_{\mathrm{noise}}=0.10$, Shannon entropy $S=6.48046$, and covariance range $[-0.0217755,0.67924]$. (c),(d) The corresponding distributions $P_{\mathrm{sur}}(s)$ of the normalized surrogate spacings. In (c), the nearly noise-free measure remains sharply concentrated near $s\simeq1$, reflecting the overly smooth deterministic envelope. In (d), moderate stochastic roughening broadens the distribution and suppresses the accumulation of near-zero spacings, producing a GOE-like level-repulsion profile. The surrogate ratio for panel (d) is defined by Eq.~\eqref{eq:surrogate_ratio} $ \langle r\rangle_{\mathrm{sur}}=0.540339$ and is kept distinct from the exact ED ratio $\langle r\rangle_{\mathrm{ED}}$.}
}		\label{noise4}
	\end{figure}

\begin{figure}[t]
\centering
\includegraphics[width=0.88\linewidth]{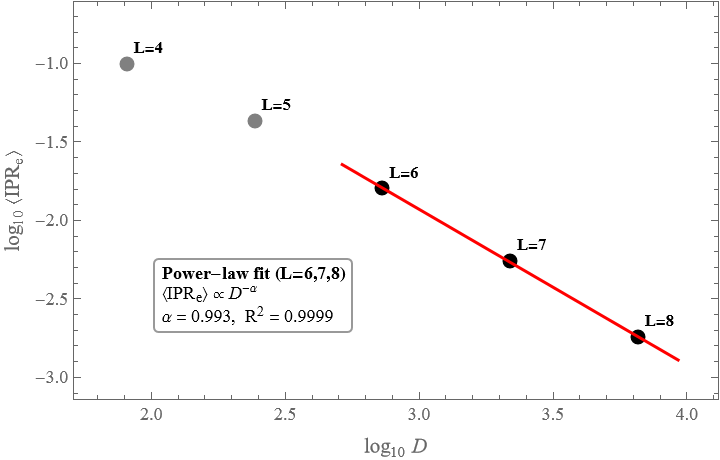}
\caption{\rev{Finite-size scaling of the exact-diagonalization eigenstate inverse participation ratio in the Fock basis. The points show $\langle\mathrm{IPR}_e\rangle$ from Table~\ref{tab:benchmark} on a $\log_{10}D$--$\log_{10}\langle\mathrm{IPR}_e\rangle$ scale for $L=4,\ldots,8$. The red line is a power-law fit to the larger-system points $L=6,7,8$, for which the adjacent-gap ratio has already stabilized near the GOE benchmark. The fit gives $\log_{10}\langle\mathrm{IPR}_e\rangle=-0.992768\,\log_{10}D+1.04891$, equivalently $\langle\mathrm{IPR}_e\rangle=11.192\,D^{-0.992768}$, with $R^2=0.999878$. The equivalent direct linear fit in $1/D$ is $\langle\mathrm{IPR}_e\rangle=11.5997/D+9.93375\times10^{-5}$ with $R^2=0.999897$. Both representations are consistent with the asymptotic $D^{-1}$ ergodic scaling; the log--log representation is shown because it provides clearer visual separation of the larger-system data and directly displays the fitted scaling exponent.}}
\label{fig:ipr_scaling}
\end{figure}

\section{\rev{Detailed Implementation of the Attention-Inspired Mechanism}}
\label{app:self_attention}

In recent neural-network quantum-state (NQS) approaches, self-attention is widely used as a trainable mechanism for learning many-body correlations. Inspired by self-attention architectures for correlated many-body problems \cite{Geier}, \rev{we employ an attention-inspired Gaussian relevance mapping to organize the mixing of Fock-state features.} However, rather than directly parameterizing the microscopic many-body wave function, 
we use an attention-inspired feature map to encode the relevance between bosonic Fock configurations and to construct a coarse-grained spectral measure.

We consider the truncated Fock basis
$\{|\mathbf n^{(i)}\rangle\}_{i=1}^{D}$,
where $D$ is the effective Hilbert-space dimension and
$|\mathbf n^{(i)}\rangle = |n_1^{(i)},n_2^{(i)},\dots,n_L^{(i)}\rangle$
denotes the $i$-th bosonic occupation configuration.
 Each configuration is represented by a local feature vector.
The input state is represented as a feature matrix $\mathbf{X} \in \mathbb{R}^{D \times d_{\mathrm{loc}}}$, where $d_{\mathrm{loc}} = N_{\max}+1$ is the local feature dimension describing site-occupation numbers for a given many-body Fock state $|\mathbf{n}^{(i)}\rangle = |n_1^{(i)}, n_2^{(i)}, \dots, n_L^{(i)}\rangle$. \rev{For comparison with generic attention notation, one may introduce Query ($\mathbf Q$), Key ($\mathbf K$), and Value ($\mathbf V$) projections,}
\begin{equation}
    \mathbf{Q} = \mathbf{X}\mathbf{W}^Q, \quad \mathbf{K} = \mathbf{X}\mathbf{W}^K, \quad \mathbf{V} = \mathbf{X}\mathbf{W}^V.
\end{equation}
\rev{Here $\mathbf W^Q$, $\mathbf W^K$, and $\mathbf W^V$ denote the projection matrices of the generic attention representation, with $\mathbf Q,\mathbf K\in\mathbb R^{D\times d_k}$ and $\mathbf V\in\mathbb R^{D\times d_v}$. They are not independently trained in the numerical implementation used for the results of this work. Instead, the physically constrained feature map collapses each Fock configuration to its diagonal interaction energy $M_i$, and the iterative optimization acts directly on the spectral weights $w_i$. The Query--Key--Value equations therefore provide a useful embedding notation rather than an additional trainable numerical branch.}

Standard transformer architectures predominantly rely on scaled dot-product attention, $\text{Softmax}(\mathbf{Q}\mathbf{K}^T/\sqrt{d_k})$. 
Due to the discrete and highly localized nature of Fock space interactions,
instead of the standard dot-product attention $\text{Softmax}(\mathbf{Q}\mathbf{K}^T / \sqrt{d_k})$, we implement a distance-based Gaussian-kernel 
score function in the projected Query-Key space
\begin{equation}
	S_{ij}=-\frac{\|\mathbf Q_i-\mathbf K_j\|^2}{2\sigma_U^2},
\end{equation}
where $\sigma_U$ sets the interaction-energy scale. The corresponding attention matrix is then defined by a row-wise softmax normalization,
\begin{equation}
	A_{ij}
	=
	\frac{\exp(S_{ij})}{\sum_{m=1}^{D}\exp(S_{im})}
	=
	\mathrm{Softmax}_j\!\left(
	-\frac{\|\mathbf Q_i-\mathbf K_j\|^2}{2\sigma_U^2}
	\right),
	\label{eq:gaussian_attention_kernel}
\end{equation}
so that the attention weights remain normalized and can be interpreted as effective relevance weights between many-body configurations. 
Expanding the squared Euclidean distance reveals that $\|\mathbf{Q}_i - \mathbf{K}_j\|^2 = \|\mathbf{Q}_i\|^2 + \|\mathbf{K}_j\|^2 - 2\mathbf{Q}_i \mathbf{K}_j^T$. Thus, the distance-based score provides a natural conceptual bridge to the quadratic thermodynamic potential $E_i\propto(M_i-\mu_t)^2$ used in Algorithm~1. \rev{The implemented feedback realizes this idea in the reduced global-context form given in Eqs.~(\ref{eq:physical_qk})--(\ref{eq:attention_weight_update}), rather than through an independently trained pairwise $\mathbf A\mathbf V$ branch.}

This Gaussian kernel measures the generalized distance between different occupation-number features in the projected Query--Key space and provides the generic pairwise form from which the reduced physical relevance mapping is motivated. \rev{For completeness, the corresponding generic attention output would be the nonlocal feature matrix $\mathbf H\in\mathbb R^{D\times d_v}$ obtained by weighting the Value matrix with the attention scores, although this $\mathbf H$ is not used as a separate optimized variable in Algorithm~1:}
\begin{equation}
    \mathbf{H} = \text{Attention}(\mathbf{Q}, \mathbf{K}, \mathbf{V}) = \mathbf{A} \mathbf{V}.
\end{equation}

\rev{Equation above states the generic pairwise attention construction. The calculations reported in this work use a reduced, physically constrained global-context form rather than evaluating $\mathbf H$ as an independently optimized feature branch. Specifically, the implemented mapping can be written as
\begin{equation}
Q_i^{\rm phys}=M_i,
\qquad
K_t^{\rm glob}=\mu_t=\sum_{j=1}^{D}w_j^{(t)}M_j,
\label{eq:physical_qk}
\end{equation}
with the sign-adaptive relevance score
\begin{equation}
S_i^{\rm glob}(t)
=\eta\frac{\Delta_{\sigma^2}}{\sigma_t^2}
\left(Q_i^{\rm phys}-K_t^{\rm glob}\right)^2
=E_i^{(t)}.
\label{eq:global_attention_score}
\end{equation}
The corresponding update is
\begin{equation}
w_i^{(t+1)}
=\frac{\exp\!\left[\ln w_i^{(t)}+S_i^{\rm glob}(t)\right]}
{\sum_j\exp\!\left[\ln w_j^{(t)}+S_j^{\rm glob}(t)\right]},
\label{eq:attention_weight_update}
\end{equation}
which has the same structural ingredients as distance-based attention---a feature-space distance, exponential scoring, and normalized competition among configurations---while operating directly on a pooled physical context. The correspondence is intentionally described as \emph{attention-inspired}, rather than as an algebraic identity with standard pairwise Transformer self-attention: $\mu_t$ is an aggregated global context rather than a separate token key, and the sign of $\Delta_{\sigma^2}$ reverses the quadratic score between the confining (cooling) and deconfining (heating) branches. In the cooling branch the score has the usual Gaussian-affinity sign, whereas the heating branch deliberately inverts that sign to implement the thermodynamic feedback.}

Equation~(\ref{eq:gaussian_attention_kernel}) may be viewed as a physically motivated generalization of the schematic attention map introduced in the main text. While the simplified form
\[
\mathrm{Softmax}\!\left(\frac{\mathbf Q\mathbf K^T}{\sigma_U}\right)
\]
captures the basic idea that attention scores quantify configuration relevance, the Gaussian kernel explicitly penalizes large separations in the projected Query-Key space and is therefore more naturally aligned with the quadratic thermodynamic modulation potential used in Algorithm~1.

The diagonal interaction energies
\begin{equation}
	M_i \equiv \langle i|\hat H_{\mathrm{int}}|i\rangle
\end{equation}
act as hidden variables defined on the Fock basis, while the off-diagonal hopping and boundary driving terms generate the effective mixing between these hidden states. The coherent driving field $F(\hat a_1^\dagger+\hat a_1)$ breaks the global $U(1)$ particle-number conservation and enables transitions between sectors with different total particle numbers. 
The nonlocal spatial correlations are governed by the dynamic competition among the boundary drive $F$, the kinetic hopping $J$, and the on-site Kerr repulsion $U$: the drive $F$ injects and extracts bosons at the boundary to couple previously disconnected particle-number sectors, the hopping $J$ transfers excitations along the 1D chain to establish spatial coherence, while the on-site interaction $U$ penalizes multiple occupancy and induces off-diagonal anti-correlations ($C_{ij} < 0$).

While in the generic pairwise formulation these configuration resonances are formally encoded by the full attention matrix $A_{ij} = \mathrm{Softmax}_j(-\|Q_i - K_j\|^2 / 2\sigma_U^2)$, our reduced numerical framework directly operationalizes this mechanism without storing a $D \times D$ attention matrix. Macroscopically, the mixing strength between different Fock sectors induced by the off-diagonal operators $K$ ($J$ and $F$) is captured through the trace invariant $\frac{1}{D}\mathrm{Tr}(K^2)$ in the target variance $\sigma_{\mathrm{target}}^2$. Microscopically, the redistribution of probability mass across Hilbert space is dynamically driven by the global-context attention score $E_i^{(t)} = \eta \frac{\Delta_{\sigma^2}}{\sigma_t^2}(M_i - \mu_t)^2$, where the local configuration energy $M_i$ acts as a query feature evaluated against the aggregated background key $\mu_t = \sum_j w_j M_j$. The subsequent normalized exponential update in Eq.\ref{14},
iteratively redistributes probability mass across the accessible multi-particle configurations. The optimized measure $\{w_i\}$ directly reconstructs the many-body particle-number covariance matrix in \ref{C1},
thereby capturing the emergent nonlocal correlations and macroscopic compressibility ($C_{ii}$) of the chaotic phase without computing microscopic wave-function phases.
\rev{In the generic pairwise construction, the attention matrix provides a relevance structure over Fock configurations. In the reduced numerical implementation, the same physical role is carried directly by the global-context score $E_i^{(t)}$ and the normalized weights $w_i$: the off-diagonal hopping and boundary driving terms motivate configuration mixing, while the diagonal $M_i$ (from $U$) and pooled mean $\mu_t$ determine the thermodynamic relevance score used in the update. No independently propagated feature matrix $\mathbf H$ is required for the reported numerical results.}

\rev{Unlike a conventional neural quantum state optimized by variational Monte Carlo (VMC), the present implementation uses the attention-inspired mapping as a physically constrained feature/relevance map and applies the thermodynamic loss directly to the reconstructed spectral weights. The optimized object is therefore the coarse-grained measure $\rho_{\mathrm{eff}}(E)=\sum_iw_i\delta(E-M_i)$, with the variance mismatch $\mathcal L=\frac12(\sigma_{\mathrm{target,eff}}^2-\sigma_t^2)^2$ driving the weight update of Algorithm~1. No independent optimization of $\mathbf W^Q$, $\mathbf W^K$, $\mathbf W^V$, or an additional MLP is used in the numerical results reported here.}
\rev{Problem-dependent weighting has also been used in variational counterdiabatic methods, where weighted actions emphasize selected matrix elements and can incorporate nonlocal information into variational coefficients\cite{Ohga2026}. The role of the weights is different here: $w_i$ are normalized probabilities over Fock configurations rather than weights in an operator action. We therefore use this comparison only as a general illustration of how targeted weighting can retain selected many-body information in a reduced description.}

Our algorithm does not optimize a full neural-network wave-function ansatz by gradient descent on a microscopic wave-function loss, but a coarse-grained surrogate representation of configuration relevance. The actual optimization target is the emergent spectral measure
$	\rho_{\mathrm{eff}}(E)=\sum_{i=1}^{D} w_i\,\delta(E-M_i)$,
whose first and second moments are constrained to match the target many-body spectrum. 
At the deterministic level (Algorithm 1), this feedback dynamics iteratively updates the weights $w_i$ to reproduce the effective target variance,
generates a smooth coarse-grained envelope for the spectral weights. 
At the stochastic level (Algorithm 2), local multiplicative fluctuations are restored to reconstruct the spacing ensemble and its Wigner-Dyson statistics. \rev{By mapping the discrete Fock-space features to a coarse-grained spectral measure, the method avoids computing the full eigenspectrum during the reconstruction; ED is retained only for the finite-size validation presented in Sec.~VI.}
\rev{Thus Algorithm~1 captures the broad macroscopic envelope, while Algorithm~2 adds phenomenological local fluctuations to the surrogate spacing measure rather than reconstructing microscopic eigenstate amplitudes.}

The final object being optimized is therefore the coarse-grained spectral distribution rather than the microscopic quantum state itself. 

	\renewcommand\refname{References}

\end{document}